\documentclass[intlimits,amssymb,12pt]{iopart}
\usepackage{bm}
\usepackage{abraces}

\usepackage{iopams}  
\usepackage{graphicx}
\begin{document}

\title[Theoretical description of pump/probe experiments in electron mediated \ldots]{Theoretical description of
pump/probe experiments in electron mediated charge-density-wave insulators}

\author{J. K. Freericks, O. Matveev, and Wen Shen}
\address{Department of Physics, Georgetown University, 37th and O Sts. NW, Washington, DC 20057 USA}
\author{A. M. Shvaika}
\address{Institute of Condensed Matter Physics of the National Academy of Sciences of Ukraine,
Lviv, 79011 Ukraine}
\author{T. P. Devereaux}
\address{Geballe Laboratory for Advanced Materials, Stanford University, 
Stanford, CA 94305,USA}
\address{Stanford Institute for Materials and Energy Sciences (SIMES), 
SLAC National Accelerator Laboratory, Menlo Park, CA 94025, USA}
\ead{james.freericks@georgetown.edu}
\vspace{10pt}
\begin{indented}
\item[]\today
\end{indented}

\begin{abstract}
In this review, we develop the formalism employed to describe charge-density-wave insulators in
pump/probe experiments using ultra short driving pulses of light. The theory emphasizes exact results
in the simplest model for a charge-density wave insulator (given by a noninteracting systems with two bands and a gap)
and by employing nonequilibrium dynamical mean-field theory to solve the Falicov-Kimball model in its ordered
phase. We show both how to develop the formalism and how the solutions behave. Care is taken to describe the details behind these calculations and to show how to verify their accuracy via sum-rule constraints.
\end{abstract}

%
\vspace{2pc}

%
\submitto{Physica Scripta}
%
%
%

\section{Introduction}

Time-resolved pump/probe experiments have seen a rebirth in recent years as timescales have been pushed well into the femtosecond (and sometimes attosecond) range, and as many different types of experiments can now be performed in an ultrafast pump/probe format. Much of the experimental work, ranging from ultrafast optical studies to angle-resolved photoemission (ARPES) have been centered on examining the behavior of charge density wave materials~\cite{perfetti1,perfetti2,rossnagel1,dragan1,dragan2,msu,shen1,shen2,rossnagel2,rossnagel3}.
In this paper, we will review work on theoretical descriptions of pump/probe
time-resolved photoemission spectroscopy on charge-density-wave (CDW) materials. Because this is a short review, we will focus entirely on our work, which describes many of the experimental features seen 
in recent studies.

Pump/probe photoemission experiments really came to life when it was recognized that one could period quadruple a 1.5~eV light pulse to create 6~eV pulses that also are ultrashort (typical widths are in the few 10s of femtoseconds). The higher-energy pulses are sufficient to photoemit electrons from part of the Brillouin zone (BZ) in materials with small workfunctions. The experimental apparatus for this was then applied to a wealth
of different materials, with a large focus on CDW systems like TaS$_2$, TiSe$_2$ and TbTe$_3$. These materials have complex ground states, and often have a number of different competing phases that can be accessed by changing temperature or pressure. 

TaS$_2$ was the first material studied~\cite{perfetti1,perfetti2}. In the 1T phase, this material goes through a range of different phase transitions, starting with an incommensurate CDW at high temperature, passing to a nearly commensurate CDW, and then a commensurate one, which has a three-sublattice ordering in the form of planar star-of-Davids. This lowest temperature commensurate CDW is predicted to be a metal in density functional theory calculations, but is seen to be an insulator in transport measurements (and other probes). It has long been believed that the insulator arises from a Mott transition in the band at the Fermi energy~\cite{tossati}, but recent work has called this interpretation into question~\cite{millis,tas2_dft}, as it is becoming clearer that the stacking of the orbital ordering along the c-axis also plays an important role, and could even be the origin of the insulating behavior.
The time-resolved ARPES studies showed that the system has its gap collapse when it is driven and it also saw an interesting subgap resonance as the system relaxed back to equilibrium. An ultrafast core-level x-ray photoemission spectroscopy study measured the charge modulation order parameter of the material~\cite{rossnagel1} and found that it decreases, and then relaxes back, but never goes all the way to zero, even though the gap has collapsed. More recently, it has been found that the system can relax to a metastable metallic nonequilibrium state~\cite{dragan1,dragan2} that appears to be similar in many respects to the nearly commensurate CDW but with some clear differences. Electron diffraction studies found yet another nonequilibrium metastable phase~\cite{msu}.

Other materials have also been studied. TbTe$_3$ is another interesting material which has been examined with time resolved ARPES~\cite{shen1,shen2}. Here one can watch the closing and re-opening of the gap directly in momentum space. In addition to seeing this behavior, two new results have emerged---first, the photoemission rings at a frequency given by the ordering phonon in the system and second, the spectral gap can be reduced, but not all the way to zero, even if there is a high fluence and there are significant subgap states which have closed the gap. The persistence of this spectral gap feature is something that naturally emerges in nesting-based CDWs as we describe below.

Finally, TiSe$_2$ has been thoroughly studied~\cite{rossnagel2,rossnagel3}. This material is believed to be an excitonic insulator, which should respond fast to a pump because the ordering is electronic in nature and does not require phonons. Indeed, the response time is seen to be quite fast in this system (under 50 femtoseconds), and a recent theory has been developed to describe the experiments~\cite{eckstein}. 

 It is clear that there is much more work needed to fully understand the behavior of these materials.

Many-body theory has also seen a significant development in recent years. Dynamical mean-field theory~\cite{metzner_vollhardt,brandt,kotliar} (DMFT) has emerged as one of the most useful methods for describing electron correlations in three-d materials. Its extension to nonequilibrium~\cite{freericks_nedmft,nedmft_review}, enables a wide range of different many-body problems to be solved. The theory can work both in the normal state and in commensurate ordered phases, and recently, there has been a range of work on the nonequilbrum properties of CDWs.  This started with the exact solution of a simplified bandstructure model, where a range of different phenomena were studied: (i) pump/probe photoemission spectroscopy~\cite{cdw_pes}; (ii) high-harmonic generation of light~\cite{cdw_hhg}; (iii) response to a large dc current and Bloch oscillations~\cite{cdw_dc}; and (iv) how quantum systems are excited as functions of the pump amplitude and frequency~\cite{cdw_excite}. Since then, this approach has been expanded to examine the behavior of nesting-driven electronic CDWs as described by the Falicov-Kimball model~\cite{falicov_kimball} and also being solved exactly~\cite{cdw_ischia,cdw_dc_fk,cdw_spie,cdw_prb}. Here, the Bloch oscillations in response to a large dc field, and the time-resolved photoemission spectroscopy were both studied. The CDW phase of the Falicov-Kimball model is quite interesting, because it displays an additional tricritical quantum-critical point, which is unique in that the order parameter is not suppressed to zero at the critical point---instead, the system transforms from an insulator to a metal~\cite{krish,cdw_transport_jkf,lemanski}. We describe this unique phenomena in detail here.

This review is not intended to be an exhaustive review of the subject, instead, its intent is to carefully derive the needed formalism to solve these problems, explain how one controls the numerics to implement the solutions, and then discusses the properties of those solutions. Our focus has been entirely on our work, which represents nearly all of the work on these systems within the nonequilibrium DMFT approach. Finally, we discuss how one can relate the results of the theory to experiment, and determine the origin of some of the features seen in these experiments.

\section{Formalism} 

Most CDW order is complex in real materials and does not simply follow the Peierls' paradigm of an AB ordered
phase~\cite{mazin}. Nevertheless, the simplest case of an AB ordered phase provides a rich playground to examine the generic properties of many CDW systems~\cite{peierls_cdw}. It also is the easiest problem to solve, and hence we focus all of our efforts in this work to understanding the properties of such an ordered system.

We envision the lattice $\Lambda$ having $|\Lambda|$ sites and being bipartite, which means it consists of two disjoint sublattices: the A sublattice and the B sublattice, which are each connected by the hopping matrix $-t_{ij}$. The hopping only connects between sites in different sublattices and for standard systems like a simple-cubic lattice, or a square lattice, the ordering wavevector is ${\bf Q}=(\pi/a,\pi/a,\ldots,\pi/a)$, in which $a$ is the lattice spacing; we work in units where $a=1$ here. This allows us to express the formalism in two representative ways---in real space or in momentum space. Due to the ordering, the translational symmetry is reduced in real space, requiring a lattice with a basis, while in momentum space, we have a coupling between momentum ${\bf k}$ and ${\bf k}+{\bf Q}$ for all momenta in the reduced Brillouin zone, which is half the size of the original BZ. The ordering wavevector satisfies
\begin{equation}
e^{i\mathbf{Q}\cdot\mathbf{R}_i}=\left \{
 1,  \quad \mathbf{R}_i\in A;\quad\quad
 -1,  \quad \mathbf{R}_i\in B, \right \},
\end{equation}
where $\mathbf{R}_i$ is the position vector of the $i$th lattice site. Note that the bandstructure for a periodic lattice,
where $-t_{ij}$ depends only on $|\mathbf{R}_i-\mathbf{R}_j|$, can be expressed as
\begin{equation}
\epsilon({\bf k})=-\sum_{j\in\Lambda} t_{ij} \rme^{i\mathbf{k}\cdot(\mathbf{R}_i-\mathbf{R}_j)}.
\label{eq: bandstructure}
\end{equation}

The Hamiltonian will be written in terms of spinless fermionic creation $c_i^\dagger$ and annihilation $c_i^{\phantom\dagger}$ operators for conduction electrons at site $i$ on the lattice.  These operators satisfy the conventional anticommutation relations $\{c_i^\dagger,c_j^{\phantom\dagger}\}_+=\delta_{ij}$ and can be transformed to momentum space via
\begin{equation}
c_{\bf k}^{\phantom\dagger}=\frac{1}{|\Lambda|}\sum_{i\in\Lambda}e^{i\mathbf{k}\cdot\mathbf{R}_i}c_i^{\phantom\dagger},\quad\quad
\textrm{and}\quad\quad c_{\bf k}^{\dagger}=\frac{1}{|\Lambda|}\sum_{i\in\Lambda}e^{-i\mathbf{k}\cdot\mathbf{R}_i}c_i^{\dagger}.
\end{equation}
In the ordered phase, the reduced Brillouin zone (rBZ) is defined via $\epsilon(\mathbf{k})\le 0$ when there is only nearest-neighbor hopping and $\mathbf{k}$ 
is restricted to the rBZ. Every momentum $\mathbf{k}$ is then coupled to the momentum $\mathbf{k}+\mathbf{Q}$ which
lies in the original BZ, but outside the rBZ.

In momentum space, the two fermionic operators are denoted $ c_{1\mathbf{k}}^{\phantom\dagger}=c_{\mathbf{k}}^{\phantom\dagger}$ and $ c_{2\mathbf{k}}^{\phantom\dagger}=c_{\mathbf{k}+\mathbf{Q}}^{\phantom\dagger}$, whereas, for the sublattice basis, we restrict the momentum summations to the respective sublattices, as follows:
\begin{equation}
c_{\mathbf{k}A}^{\phantom\dagger}=\frac{\sqrt{2}}{|\Lambda|}\sum_{i\in A}e^{i\mathbf{k}\cdot\mathbf{R}_i}c_i^{\phantom\dagger}\quad\quad\textrm{and}\quad\quad
c_{\mathbf{k}B}^{\phantom\dagger}=\frac{\sqrt{2}}{|\Lambda|}\sum_{i\in B}e^{i\mathbf{k}\cdot\mathbf{R}_i}c_i^{\phantom\dagger},
\end{equation}
where the $\sqrt{2}$ is required so that these fermionic operators satisfy the conventional anticommutation relations. With this notation, we then find that
\begin{equation}
c_{1\mathbf{k}}^{\phantom\dagger}=\frac{c_{\mathbf{k}A}^{\phantom\dagger}+c_{\mathbf{k}B}^{\phantom\dagger}}{\sqrt{2}}\quad\quad\textrm{and}\quad\quad
c_{2\mathbf{k}}^{\phantom\dagger}=\frac{c_{\mathbf{k}A}^{\phantom\dagger}-c_{\mathbf{k}B}^{\phantom\dagger}}{\sqrt{2}}.
\end{equation}
As we work on the development of the many-body formalism for pump/probe experiments on CDW systems, we will find it convenient to use both of these different bases to represent the fermionic creation and annihilation operators. Converting between the two is simple, with the unitary transformation that relates them satisfying
\begin{equation}
\left (
   \begin{array}{c}
  {c}_{1\mathbf{k}}^{\phantom\dagger} \\
  {c}_{2\mathbf{k}}^{\phantom\dagger}
   \end{array}
\right )
=\left (\begin{array}{c c}
  \frac{1}{\sqrt{2}} &  \frac{1}{\sqrt{2}} \\
  \frac{1}{\sqrt{2}} & -\frac{1}{\sqrt{2}}
   \end{array}
\right )\left ( \begin{array}{c}
  c_{\mathbf{k} A}^{\phantom\dagger} \\
  c_{\mathbf{k} B}^{\phantom\dagger}
   \end{array}
\right ) .
\end{equation}

We will be working with an all electronic model of the CDW, which can be described by the Falicov-Kimball model 
(or a specific, noninteracting limit of the model described below)~\cite{falicov_kimball}, which has conduction electrons, denoted with a $c$
and localized electrons, denoted by an $f$, which mutually interact. The Hamiltonian in an external electric field is
\begin{equation}
\mathcal{H}^{FK}(t)=-\sum_{ij\in\Lambda}t_{ij}(t)c^\dagger_ic^{\phantom\dagger}_j-\mu\sum_{i\in\Lambda}
c^\dagger_ic^{\phantom\dagger}_i+E_f\sum_{i\in\Lambda}w_i+U\sum_{i\in\Lambda}c^\dagger_ic^{\phantom\dagger}_iw_i,
\label{eq: ham}
\end{equation}
where $w_i=f^\dagger_if^{\phantom\dagger}_i=0,~1$ is the localized electron number operator (which can be treated
as a classical Ising-like variable), $\mu$ is the conduction electron chemical potential, $E_f$ is the $f$-electron site energy
and $U$ is the on-site Coulomb repulsion. 

In the limit $T\rightarrow 0$, and in the case of half-filling, where the
density of conduction electrons and the density of localized electrons is each 0.5, we find that the equilibrium
solution has $w_i=1$ for $i\in A$ and $w_i=0$ for $i\in B$. The simplest model for a CDW fixes $w_i$ at those
exact values for all $T$. For the simplified model, the system always has CDW order, and it can be solved by diagonalizing a bandstructure on a lattice with a basis (although the dynamics are still complex in nonequilibrium). While in the Falicov-Kimball model, the asymmetry between the two sublattices decreases 
as the temperature increases until we reach $T_c$, where the density of both particles becomes uniformly distributed, on average, throughout the entire lattice. In Eq.~(\ref{eq: ham}), the hopping matrix is time-dependent to model an electric field via the Peierls substitution~\cite{peierls_subst}
\begin{equation}
-t_{ij}(t)=-t_{ij}\rme^{i\mathbf{A}(t)\cdot(\mathbf{R}_i-\mathbf{R}_j)},
\label{eq: hopping}
\end{equation}
with the spatially uniform, but time-dependent electric field given by the negative of the time derivative of the spatially uniform vector potential: $\mathbf{E}(t)=-d\mathbf{A}(t)/dt$. In Eq.~(\ref{eq: hopping}), the hopping matrix $-t_{ij}$
is a constant spatially periodic matrix [which we take to be nonzero only between nearest neighbors, where it is equal to
$t^*/(2\sqrt{d})$, with $d$ the spatial dimension of the system; we will work in units with $t^*=1$ and take the $d\rightarrow\infty$ limit]. In other words, we will be working on the infinite-dimensional hypercubic lattice.

We re-express the kinetic-energy operator in terms of the bandstructure in Eq.~(\ref{eq: bandstructure}) and the conduction-electron operators in the two different representations (when in the ordered phase):
\begin{eqnarray}
\fl
\quad\quad
-\sum_{ij\in\Lambda}t_{ij}(t)c^\dagger_ic^{\phantom\dagger}_j&=&\sum_{\mathbf{k}\in \textrm{\footnotesize rBZ}}
\aoverbrace[L1R]{\aunderbrace[l1r]{c_{\mathbf{k}A}^\dagger~c_{\mathbf{k}B}^\dagger}}
\left ( \begin{array}{c c}
0 & \epsilon(\mathbf{k}-\mathbf{A}(t))\\
 \epsilon(\mathbf{k}-\mathbf{A}(t)) & 0
\end{array}
\right )
\left (
\begin{array}{c}
c_{\mathbf{k}A}^{\phantom\dagger}\\
c_{\mathbf{k}B}^{\phantom\dagger}
\end{array}
\right ),
\label{eq: ke_ab}\\
&=&\sum_{\mathbf{k}\in \textrm{\footnotesize rBZ}}
\aoverbrace[L1R]{\aunderbrace[l1r]{c_{\mathbf{k}1}^\dagger~c_{\mathbf{k}2}^\dagger}}
\left ( \begin{array}{c c}
\epsilon(\mathbf{k}-\mathbf{A}(t))&0\\
 0&-\epsilon(\mathbf{k}-\mathbf{A}(t))
\end{array}
\right )
\left (
\begin{array}{c}
c_{\mathbf{k}1}^{\phantom\dagger}\\
c_{\mathbf{k}2}^{\phantom\dagger}
\end{array}
\right ).
\label{eq: ke_12}
\end{eqnarray}
Here, the bandstructure shifted by the vector potential is directed along the diagonal of the hypercubic lattice and thereby 
satisfies
\begin{equation}
\epsilon(\mathbf{k}-\mathbf{A}(t))=\epsilon(\mathbf{k})\cos{A}(t)+\tilde\epsilon(\mathbf{k})\sin{A}(t),
\end{equation}
with
\begin{equation}
\fl
\quad\quad
\epsilon(\mathbf{k})=\lim_{d\rightarrow\infty}\left (-\frac{t^*}{\sqrt{d}}\sum_{i=1}^d\cos {k}_i\right )\quad\textrm{and}\quad
\tilde\epsilon(\mathbf{k})=\lim_{d\rightarrow\infty}\left (-\frac{t^*}{\sqrt{d}}\sum_{i=1}^d\sin {k}_i\right )
\end{equation}
and the field (and vector potential) oriented along the diagonal so that $\mathbf{A}(t)=(A(t),A(t),\cdots,A(t))$.
One can think of the second bandstructure $\tilde\epsilon$ as the projection of the velocity onto the direction of the 
electric field.
Note how the $2\times 2$ matrix that represents the kinetic energy is off-diagonal in the sublattice representation and diagonal in the original momentum representation ($\mathbf{k}$ and $\mathbf{k}+\mathbf{Q}$).

\begin{figure}
\centerline{
\includegraphics[width=0.48\textwidth]{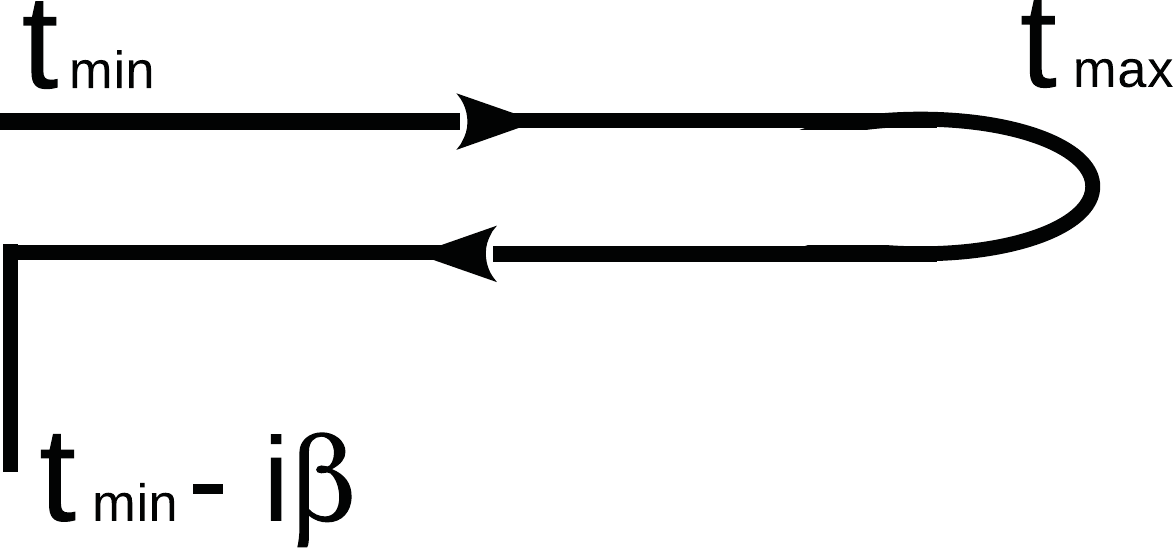}}
 \caption{Kadanoff-Baym-Keldysh contour which runs from a minimum time $t_{\rm min}$ to a maximum time $t_{\rm max}$ and back, ending with a spur parallel to the negative imaginary axis of length $\beta=1/T$.}
 \label{fig: keldysh}
\end{figure}

The many-body problem is solved with contour-ordered Green's functions which are defined on the 
Kadanoff-Baym-Keldysh contour depicted in Fig.~\ref{fig: keldysh}~\cite{kadanoff_baym,keldysh}. The contour graphically illustrates the 
time evolution of the operators in the Heisenberg representation, which evolve from the intial time ($-\infty$) to time $t$, then from time $t$ to $t'$ and finally from time $t'$ back to the initial time, followed by an evolution along a segment of length $\beta=1/T$
parallel to the negative imaginary axis. The momentum-dependent contour-ordered Green's function (for $\mathbf{k}$ in the rBZ) is defined by
\begin{equation}
G^c_{\mathbf{k}rs}(t,t')=-i\Tr \mathcal{T}_c\frac{\rme^{-\beta\mathcal{H}(-\infty)}}{\mathcal{Z}}c_{\mathbf{k}r}^{\phantom\dagger}(t)c_{\mathbf{k}s}^\dagger(t'),
\end{equation}
with $r$ and $s$ being the 1, 2 or A, B subscripts depending upon the representation and $\mathcal{Z}=\Tr \exp[-\beta\mathcal{H}(t=-\infty)]$ being the partition function. The symbol $\mathcal{T}_c$ is the time-ordering operator,
which orders times \textit{along the contour}. The fermionic creation and annihilation operators are in the Heisenberg representation where $\mathcal{O}(t)=\mathcal{U}^\dagger(t,-\infty)\mathcal{O}\mathcal{U}(t,-\infty)$ and the evolution operator satisfies
\begin{equation}
\mathcal{U}(t,-\infty)=\mathcal{T}_t \rme^{-i\int_{-\infty}^t d\bar t\mathcal{H}(\bar t)},
\end{equation}
where the time ordering is with respect to ordinary time. Substituting the evolution operators into the definition of the Green's function then yields
\begin{equation}
G^c_{\mathbf{k}rs}(t,t')=-i\Tr \mathcal{T}_c\frac{\rme^{-\beta\mathcal{H}(-\infty)}}{\mathcal{Z}}
\mathcal{U}(-\infty,t)c_{\mathbf{k}r}^{\phantom\dagger}\mathcal{U}(t,t')c_{\mathbf{k}s}^\dagger
\mathcal{U}(t',-\infty),
\end{equation}
for $t$ ahead of $t'$ on the contour ($t>_ct'$); we employed the identities $\mathcal{U}^\dagger(t,t')=\mathcal{U}(t',t)$ and $\mathcal{U}(t,t'')\mathcal{U}(t'',t')=\mathcal{U}(t,t')$. One can now directly see why the contour runs from the initial time to $t$, back to the initial time, and then along the imaginary axis (if we think of the thermal factor as an evolution
along the imaginary-time axis).

There are two Green's functions that we can extract from the contour-ordered Green's functions---the retarded Green's function (which holds information about the quantum states) and the lesser Green's function (which tells us how those states are occupied). They are defined via
\begin{equation}
G^R_{\mathbf{k}rs}(t,t')=-i\theta(t-t')\Tr \frac{\rme^{-\beta\mathcal{H}(-\infty)}}{\mathcal{Z}}\{c_{\mathbf{k} r}^{\phantom\dagger}(t),c_{\mathbf{k}s}^\dagger(t')\}_+,
\end{equation}
and
\begin{equation}
G^<_{\mathbf{k}rs}(t,t')=-i\Tr \frac{\rme^{-\beta\mathcal{H}(-\infty)}}{\mathcal{Z}}c_{\mathbf{k}s}^\dagger(t')c_{\mathbf{k} r}^{\phantom\dagger}(t).
\end{equation}

The self-energy is defined via the equation of motion. To begin, we must first determine the noninteracting Green's function. This is found by setting $U=0$ in the Hamiltonian. Because the subsequent Hamiltonian commutes with itself at different times, the Green's function can be found simply by determining the equation of motion for the fermionic creation and annihilation operators. This yields (with the integrals between $t'$ and $t$ \textit{on the contour})
\begin{eqnarray}
\fl
~~~&G^{c,{\scriptstyle\rm nonint}}_{\mathbf{k}}(t,t')=\nonumber\\
\fl ~~~&i\left ( 
\begin{array}{c c}
{\scriptstyle [f(\epsilon(\mathbf{k})-\mu)-\theta_c(t,t')]}\rme^{-i\int_{t'}^td\bar t [\epsilon(\mathbf{k}-\mathbf{A}(t))-\mu]} & 0\\
\fl
0 & {\scriptstyle [f(-\epsilon(\mathbf{k})-\mu)-\theta_c(t,t')]}\rme^{-i\int_{t'}^td\bar t [-\epsilon(\mathbf{k}-\mathbf{A}(t))-\mu]}
\end{array}
\right ),
\end{eqnarray}
since $\epsilon(\mathbf{k}+\mathbf{Q})=-\epsilon(\mathbf{k})$. We introduced the Fermi-Dirac distribution
$f(x)=1/[1+\exp(\beta x)]$ and the contour unit step function $\theta_c(t,t')$ which is equal to 1 if $t>_ct'$ and 0 if $t<_c t'$. This is in the $1,2$ momentum representation, where the kinetic energy is diagonal.

The self-energy is diagonal in the $A,B$ sublattice representation, given by diagonal elements $\Sigma^c_A(t,t')$ and $\Sigma^c_B(t,t')$;
it has no momentum dependence because we are solving the problem in dynamical mean-field theory, which has a local self-energy. Converting to the $1,2$ representation yields
\begin{equation}
\Sigma^c(t,t')=\left (
\begin{array}{c c}
\frac{1}{2}[\Sigma^c_A(t,t')+\Sigma^c_B(t,t')] & \frac{1}{2}[\Sigma^c_A(t,t')-\Sigma^c_B(t,t')]\\
\frac{1}{2}[\Sigma^c_A(t,t')-\Sigma^c_B(t,t')] & \frac{1}{2}[\Sigma^c_A(t,t')+\Sigma^c_B(t,t')]
\end{array}
\right ).
\end{equation}
Dyson's equation then yields the interacting Green's function
\begin{equation}
G^c_{\mathbf{k}}(t,t')=\left [ \left (G^{c,{\scriptstyle\rm nonint}}_{\mathbf{k}}\right )^{-1}-\Sigma^c\right ]^{-1}(t,t'),
\label{eq: dyson}
\end{equation}
which is the $t,t'$ matrix element of the inverse of the operator inside the square brackets.  The inverse is with respect to both the time indices and the $2\times 2$ structure imposed by the ordered phase. Note that the Green's functions and the self-energy are continuous matrix operators. Hence, they cannot be easily evaluated numerically. To do so, requires one
to discretize the contour and approximate the operators by finite matrices, which can then be inverted using standard linear algebra methods. The discretization is then extrapolated to zero to produce the approximation to the continuous matrix operator. Details for how to do this can be found in Ref.~\cite{freericks_nedmft2}.

We are often interested in local quantities, such as the local Green's function, which is found by summing the momentum-dependent Green's function over all momentum in the rBZ. In many cases, it is more convenient to perform the summation over the entire BZ, since each term appears twice in the summation and one does not need to trace over the final $2\times 2$ structure. The difference is whether one sums over $\epsilon$ terms that are larger than zero.  If restricting to the rBZ, then we must weight the $\epsilon=0$ terms by 0.5, otherwise, those boundary terms will be over counted. Of course, because the self-energy is independent of momentum, the sum over momentum can be replaced by a two-dimensional integral over the joint density of states for $\epsilon$ and $\tilde\epsilon$. In the $d\rightarrow\infty$ limit, we find that the joint density of states is a double Gaussian given by~\cite{nedmft_nonint}
\begin{equation}
\rho(\epsilon,\tilde\epsilon)=\frac{1}{\pi}\rme^{-\epsilon^2-\tilde\epsilon^2}.
\end{equation}
If restricting to the rBZ, then one needs to reweight the joint density of states, and in some cases perform a trace over the final $2\times 2$ matrix. But, by convention, we often weight the Green's functions so that the many-body density of states on the A sublattice and the B sublattice each have spectral weight of 1. In that case, one averages over both to get the \textit{average} local density of states.

The simplified CDW case is determined more easily than the full interacting case.  The self-energy simplifies to $\Sigma^c_A=U$ and $\Sigma^c_B=0$. Then, the evolution operator becomes block-diagonal for each momentum
(or, equivalently for each $\epsilon$, $\tilde\epsilon$ pair), which is described by a simple $2\times 2$ Landau-Zener-like system. The full evolution operator is found by using the Trotter formula for a given discretization $\Delta t$ and can
be analytically found to satisfy
\begin{eqnarray}
\mathcal{U}(\mathbf{k},t+\Delta t,t)=\cos\left [ \Delta t\sqrt{\epsilon^2\left ({\bf k}-{\bf A}\left (t+\frac{\Delta t}{2}\right )\right )+\frac{U^2}{4}}\right ] \mathbb{I}\\
-i\left [\epsilon\left ( \mathbf{k}-\mathbf{A}\left ( t+\frac{\Delta t}{2}\right )\right )\sigma_z+\frac{U}{2}\sigma_x\right ]\frac{\sin\left [ \Delta t\sqrt{\epsilon^2\left ( \mathbf{k}-\mathbf{A}\left ( t+\frac{\Delta t}{2}\right )\right )+\frac{U^2}{4}}\right ]}{\sqrt{\epsilon^2\left  (\mathbf{k}-\mathbf{A}\left ( t+\frac{\Delta t}{2}\right )\right )+\frac{U^2}{4}}},
\nonumber
\end{eqnarray}
where $\mathbb{I}$ is the $2\times 2$ identity matrix, $\sigma_x$ and $\sigma_y$ are the corresponding Pauli spin matrices, and we employ a midpoint integration rule for the evaluation of the Hamiltonian in the Trotter factor. The full evolution operator for this $2\times 2$ block then becomes
\begin{equation}
\fl
\mathcal{U}(\mathbf{k},t,t')=\mathcal{U}(\mathbf{k},t,t-\Delta t)\mathcal{U}(\mathbf{k},t-\Delta t,t-2\Delta t)\cdots
\mathcal{U}(\mathbf{k},t'+2\Delta t,t'+\Delta t)\mathcal{U}(\mathbf{k},t'+\Delta t,t'),
\end{equation}
and then this is repeated for each momentum point.

It turns out that the Green's function for the simplified CDW model is determined entirely in terms of this evolution
operator. Namely, we find that the retarded Green's function in the 1,2 momentum representation is determined by just evolution operators
between times $t$ and $t'$ (because the quantum states just depend on the instantaneous value of the Hamiltonian) via
\begin{equation}
G^R_{\mathbf{k}}(t,t')=-i\theta(t-t')\mathcal{U}(\mathbf{k},t,t').
\end{equation}
If we sum over momentum and convert to the $A,B$ representation, we find that
\begin{equation}
\fl
G^R_{AA,BB}(t,t')=-i\theta(t-t')\sum_{\mathbf{k}}\left [\mathcal{U}_{11}(\mathbf{k},t,t')
+\mathcal{U}_{22}(\mathbf{k},t,t')\pm \mathcal{U}_{12}(\mathbf{k},t,t')\pm\mathcal{U}_{21}(\mathbf{k},t,t')\right ] ,
\end{equation}
with the $+$ sign for the $A$ sublattice and the $-$ sign for the $B$ sublattice; these Green's functions are normalized so that $G^R_{AA}(t,t)=G^R_{BB}(t,t)=1$. Note that the sum over momentum is replaced by a double integral over the two band energies weighted by the joint density of states. The lesser Green's functions are more complicated, because they depend on all times, not just the times between $t$ and $t'$. This is reasonable, because how the states are occupied depends on the history of how the occupancy has evolved over time. The final expression depends on the initial occupancies of the electrons. These are given by the following
\begin{eqnarray}
\langle c^\dagger_{\mathbf{k}}(-\infty)c^{\phantom\dagger}_{\mathbf{k}}(-\infty)\rangle = \beta_{\mathbf{k}}^2,\quad
\langle c^\dagger_{\mathbf{k}}(-\infty)c^{\phantom\dagger}_{\mathbf{k}+\mathbf{Q}}(-\infty)\rangle = -\alpha_{\mathbf{k}}\beta_{\mathbf{k}}\\
\langle c^\dagger_{\mathbf{k}+\mathbf{Q}}(-\infty)c^{\phantom\dagger}_{\mathbf{k}}(-\infty)\rangle=-\alpha_{\mathbf{k}}\beta_{\mathbf{k}},\quad
\langle c^\dagger_{\mathbf{k}+\mathbf{Q}}(-\infty)c^{\phantom\dagger}_{\mathbf{k}+\mathbf{Q}}(-\infty)\rangle=\alpha_{\mathbf{k}}^2\\
\alpha_{\mathbf{k}}=\frac{\frac{U}{2}}{\sqrt{2\left ( \epsilon^2(\mathbf{k})+\frac{U^2}{4}-\epsilon(\mathbf{k})
\sqrt{\epsilon^2(\mathbf{k})+\frac{U^2}{4}}\right )}},\\
\beta_{\mathbf{k}}=\frac{ -\epsilon(\mathbf{k})+
\sqrt{\epsilon^2(\mathbf{k})+\frac{U^2}{4}}}{\sqrt{2\left ( \epsilon^2(\mathbf{k})+\frac{U^2}{4}-\epsilon(\mathbf{k})
\sqrt{\epsilon^2(\mathbf{k})+\frac{U^2}{4}}\right )}}.
\end{eqnarray}
Here, we use the following notation $\langle \mathcal{O}\rangle=\Tr \exp(-\beta\mathcal{H}(-\infty))\mathcal{O}/\mathcal{Z}$.
The final result for the lesser Green's function in the 1,2 representation is cumbersome and is given by
\begin{eqnarray}
G_{\mathbf{k}11}^<(t,t')&=i\left [ \mathcal{U}_{11}(\mathbf{k},t')\mathcal{U}_{11}(\mathbf{k},t)\beta^2_{\mathbf{k}}
-\mathcal{U}_{11}(\mathbf{k},t')\mathcal{U}_{12}(\mathbf{k},t)\alpha_{\mathbf{k}}\beta_{\mathbf{k}}\right .\nonumber\\
&\left .-\mathcal{U}_{21}(\mathbf{k},t')\mathcal{U}_{11}(\mathbf{k},t)\alpha_{\mathbf{k}}\beta_{\mathbf{k}}+
\mathcal{U}_{21}(\mathbf{k},t')\mathcal{U}_{12}(\mathbf{k},t)\alpha_{\mathbf{k}}^2\right ],
\end{eqnarray}
\begin{eqnarray}
G_{\mathbf{k}12}^<(t,t')&=i\left [ \mathcal{U}_{11}(\mathbf{k},t')\mathcal{U}_{21}(\mathbf{k},t)\beta^2_{\mathbf{k}}
-\mathcal{U}_{11}(\mathbf{k},t')\mathcal{U}_{22}(\mathbf{k},t)\alpha_{\mathbf{k}}\beta_{\mathbf{k}}\right .\nonumber\\
&\left .-\mathcal{U}_{21}(\mathbf{k},t')\mathcal{U}_{21}(\mathbf{k},t)\alpha_{\mathbf{k}}\beta_{\mathbf{k}}+
\mathcal{U}_{21}(\mathbf{k},t')\mathcal{U}_{22}(\mathbf{k},t)\alpha_{\mathbf{k}}^2\right ],
\end{eqnarray}
\begin{eqnarray}
G_{\mathbf{k}21}^<(t,t')&=i\left [ \mathcal{U}_{12}(\mathbf{k},t')\mathcal{U}_{11}(\mathbf{k},t)\beta^2_{\mathbf{k}}
-\mathcal{U}_{22}(\mathbf{k},t')\mathcal{U}_{12}(\mathbf{k},t)\alpha_{\mathbf{k}}\beta_{\mathbf{k}}\right .\nonumber\\
&\left .-\mathcal{U}_{12}(\mathbf{k},t')\mathcal{U}_{11}(\mathbf{k},t)\alpha_{\mathbf{k}}\beta_{\mathbf{k}}+
\mathcal{U}_{22}(\mathbf{k},t')\mathcal{U}_{12}(\mathbf{k},t)\alpha_{\mathbf{k}}^2\right ],
\end{eqnarray}
and
\begin{eqnarray}
G_{\mathbf{k}22}^<(t,t')&=i\left [ \mathcal{U}_{12}(\mathbf{k},t')\mathcal{U}_{21}(\mathbf{k},t)\beta^2_{\mathbf{k}}
-\mathcal{U}_{22}(\mathbf{k},t')\mathcal{U}_{22}(\mathbf{k},t)\alpha_{\mathbf{k}}\beta_{\mathbf{k}}\right .\nonumber\\
&\left .-\mathcal{U}_{12}(\mathbf{k},t')\mathcal{U}_{21}(\mathbf{k},t)\alpha_{\mathbf{k}}\beta_{\mathbf{k}}+
\mathcal{U}_{22}(\mathbf{k},t')\mathcal{U}_{22}(\mathbf{k},t)\alpha_{\mathbf{k}}^2\right ].
\end{eqnarray}
The shortened symbol $\mathcal{U}_{rs}(\mathbf{k},t)=\mathcal{U}_{rs}(\mathbf{k},t,-\infty)$ is employed in these equations.
While the transformation to the $A,B$ representation is straightforward, the resulting equations are so long, that we do not write them down here. 

However, it is important to calculate the order parameter of the conduction electrons, for both the Falicov-Kimball model and the simplified model. It is given by
\begin{equation}
\Delta n_c(t)=\frac{n_B(t)-n_A(t)}{2[n_B(t)+n_A(t)]}=-\frac{\sum_{\mathbf{k}\in {\rm rBZ}}\left [ G^<_{12}(\mathbf{k},t,t)+G^<_{21}(\mathbf{k},t,t)\right ]}
{2\sum_{\mathbf{k}\in {\rm rBZ}}\left [ G_{11}^<(\mathbf{k},t,t)+G_{22}^<(\mathbf{k},t,t)\right ]},
\end{equation}
which is bounded between 0 and 0.5 in equilibrium (but can become negative in nonequilibrium). Similarly, the order parameter of the localized electrons is 
\begin{equation}
\Delta n_f=\frac{\langle w_{i\in A}\rangle -\langle w_{i\in B}\rangle}{2(\langle w_{i\in A}\rangle +\langle w_{i\in B}\rangle)},
\end{equation}
which is fixed at 0.5 for the simplified model, and reaches 0.5 at $T=0$ for the Falicov-Kimball model. It is always nonegative, because it is fixed at its equilibrium value, and hence also has no time dependence.

While we have provided a complete solution of the simplified model, we have not yet described how one solves the
DMFT for the CDW state in nonequilibrium. It is solved via an iterative algorithm, but we must work with
matrices that have time discretized on the contour. The algorithm starts with a guess for the self-energies on the
two sublattices (in the $A,B$ representation, the self-energy is diagonal in the $rs$ space). The iterative approach is then as follows:
(1) for the given self-energies, compute the local Green's function $G^c_{\rm local}(t,t')=\sum_{\mathbf{k}\in{\rm rBZ}}G^c_{\mathbf{k}}(t,t')$ by summing Dyson's equation in Eq.~(\ref{eq: dyson}) over all momenta (practically speaking, we use an integration over the joint density of states to do this and we do so in the $A,B$ representation);
(2) extract the effective medium $G^c_0$ from the local Dyson's equation (which has an additional $2\times 2$ matrix structure in the $A,B$ representation):
\begin{equation}
\left ( G^c_0 \right )^{-1}(t,t')=\left ( G^c_{\rm local} \right )^{-1}(t,t')+\Sigma^c(t,t');
\end{equation}
(3) construct the (diagonal) impurity Green's function from the diagonal components of the effective medium via
\begin{equation}
G^c_{\rm impurity \it rr}(t,t')=(1-n^f_r)G^c_{0rr}(t,t')+n^f_r\left [ \left (1-G^c_{0rr}U\right )^{-1} G^c_{0rr}\right ](t,t'),
\end{equation}
with $r=A$ or $B$ and $n^f_r=\langle w_i\rangle$ with $i\in r=A$ or $B$;
(4) extract the self-energy for the impurity by solving 
\begin{equation}
\Sigma^c_r(t,t')=\left ( G^c_{0 \it rr}\right )^{-1}(t,t')-\left ( G^c_{\rm impurity \it rr}\right )^{-1}(t,t');
\end{equation}
(5) and finally setting the new self-energy for the lattice to equal that of the impurity and then iterates steps (1)--(4) until it is converged.

One of the important checks is the short time behavior of the Green's function. It turns out that by carefully examining the definition of the Green's function, one can find the coefficients of the Taylor series expansion in relative time. In particular, for the retarded Green's function, when $t=t'$ (or $t_{\rm rel}=t-t'=0^+$), we immediately know that $G^R(t^+,t)=-i$,
because the anticommutator of two fermionic operators is equal to one; this holds both in momentum space and in real space. Higher derivatives can be evaluated by taking commutators with the Hamiltonian. Remarkably, the first few derivatives do not depend on the field, so they hold in equilibrium and in nonequilibrium. These results are also called moment sum rules, because that is what they look like when one converts the relative time to a frequency via Fourier transformation. Since the first few moment sum rules (or equivalently relative time derivatives) of the Green's function can be found exactly, they become an important tool in testing the accuracy of calculations~\cite{sum_rules}. For the simplified CDW model, one can evaluate a Taylor series expansion of the time evolution operator and immediately find that the expressions given above satisfy the appropriate sum rules~\cite{cdw_dc}. For the Falicov-Kimball model, we calculate the moment sum rules by numerically evaluating the first few derivatives of the retarded Green's functions. They serve as important numerical consistency checks of the approach, and are critical to ensure accuracy of the final results.

The derivation of these sum rules is straightforward, but tedious.  While one can do this both for the momentum-dependent Green's functions and the local Green's functions, we report them only for the local Green's functions here.  The standard way to report them is in terms of the many-body density of states for each sublattice, which is defined to be $A_\alpha(t_{\rm ave},\omega)=-\textrm{Im}G^R_\alpha(t_{\rm ave},\omega)$, with $G^R_\alpha(t_{ave},\omega)=\int dt_{\rm rel}G^R_\alpha(t_{\rm ave},t_{\rm rel})\exp(-i\omega t_{\rm rel})$. Here, the average time is $t_{\rm ave}=(t+t')/2$ and the relative time is $t_{\rm rel}=t-t'$. The moments then satisfy $\mu^R_{\alpha,n}=\int d\omega \omega^nA^R_\alpha(\omega)$ which become
\begin{equation}
\mu^R_{\alpha,0}=1,
\end{equation}
\begin{equation}
\mu^R_{\alpha,1}=-\mu+U\langle w_{i\in\alpha}\rangle,
\end{equation}
and
\begin{equation}
\mu^R_{\alpha,2}=\frac{1}{2}+\mu^2-2\mu U\langle w_{i\in\alpha}\rangle +U^2\langle w_{i\in\alpha}\rangle.
\end{equation}
The expectation value $\langle w_{i\in\alpha}\rangle$ measures the average density of the heavy particles on each sublattice; this expectation value does not change with time.

When calculating momentum-dependent quantities, like the angle-resolved PES, one must be careful to work with gauge-invariant quantities to ensure that the object being measured is a true observable. In this work, we will focus on the angle-summed PES, or total PES, which, being a local quantity, is manifestly gauge invariant. Hence, we do not discuss gauge-invariance issues further here. 

In addition, we work with a constant matrix element approximation. For a single band model in the normal state, a constant matrix element simply factors out of the PES expressions. But when the system has multiple bands, so that the Green's function is represented by a matrix (here a $2\times 2$ matrix structure for the $A$, $B$ sublattices), then the matrix elements cannot be constant in all different bases---they are constant in one basis, then they are related by a unitary transformation in another basis. While it is tempting to ignore this fact, and approximate the PES signal by the trace of the matrix Green's function (since it is an invariant) multiplied by a constant, this only holds in the basis where the Green's function matrix is diagonal; the photoemission spectra may not satisfy positivity in this case. Otherwise, the PES (and especially the angle-resolved PES) can involve more complex contributions. Here we take the simplifying assumption that the PES is given by a constant matrix element multiplied by the sum of the local diagonal contributions in the $A$, $B$ representation. The averaging over the two sublattices is required because the probe pulse will uniformly irradiate both sublattices. The result for the photoemission from each sublattice is then a two-time, probe-pulse-envelope-weighted, Fourier transform, given by~\cite{tr_pes}
\begin{equation}
P(\omega,t_0')=-i\int_{-\infty}^{\infty}dt\int_{-\infty}^{\infty}dt's(t)s(t')e^{-i\omega(t-t')}\sum_{\alpha=A,B}G^<_\alpha(t,t').
\label{eq: pes}
\end{equation}
The symbol $s(t)$ is the probe pulse envelope function, which we take to be a Gaussian centered about the time $t_0'$:
\begin{equation}
s(t)=\frac{1}{\sigma_b\sqrt{\pi}}e^{(t-t_0')^2/\sigma_b^2},
\label{eq: probe}
\end{equation}
with $\sigma_b$ the effective width of the probe pulse (broader pulses mean more energy resolution, less time resolution and \textit{vice versa}).

For completeness, we discuss two other observables. One is the current, which determines the time rate of change of the energy via $-\mathbf{E}\cdot\langle \mathbf{j}(t)\rangle$ and the other is the filling within each of the bands (essentially the filling within bands with energy larger than zero or smaller than zero). For the current, one finds
\begin{equation}
\langle\mathbf{j}(t)\rangle=-i\sum_{\mathbf{k}\in \rm rBZ}\mathbf{v}[\mathbf{k}-\mathbf{A}(t)]\left [ G^<_{\mathbf{k}11}(t,t)+G^<_{\mathbf{k}22}(t,t)\right ],
\end{equation}
where $\mathbf{v}=\nabla_{\mathbf{k}}\epsilon(\mathbf{k})$ is the particle band velocity and the Green's function is in the 1,2 basis. The 
result for the filling into the different bands has only been derived for the simplified model. We refer to Ref.~\cite{cdw_dc}
for those complete formulas.

\begin{figure}[htb]
\centerline{
\includegraphics[width=0.6\textwidth]{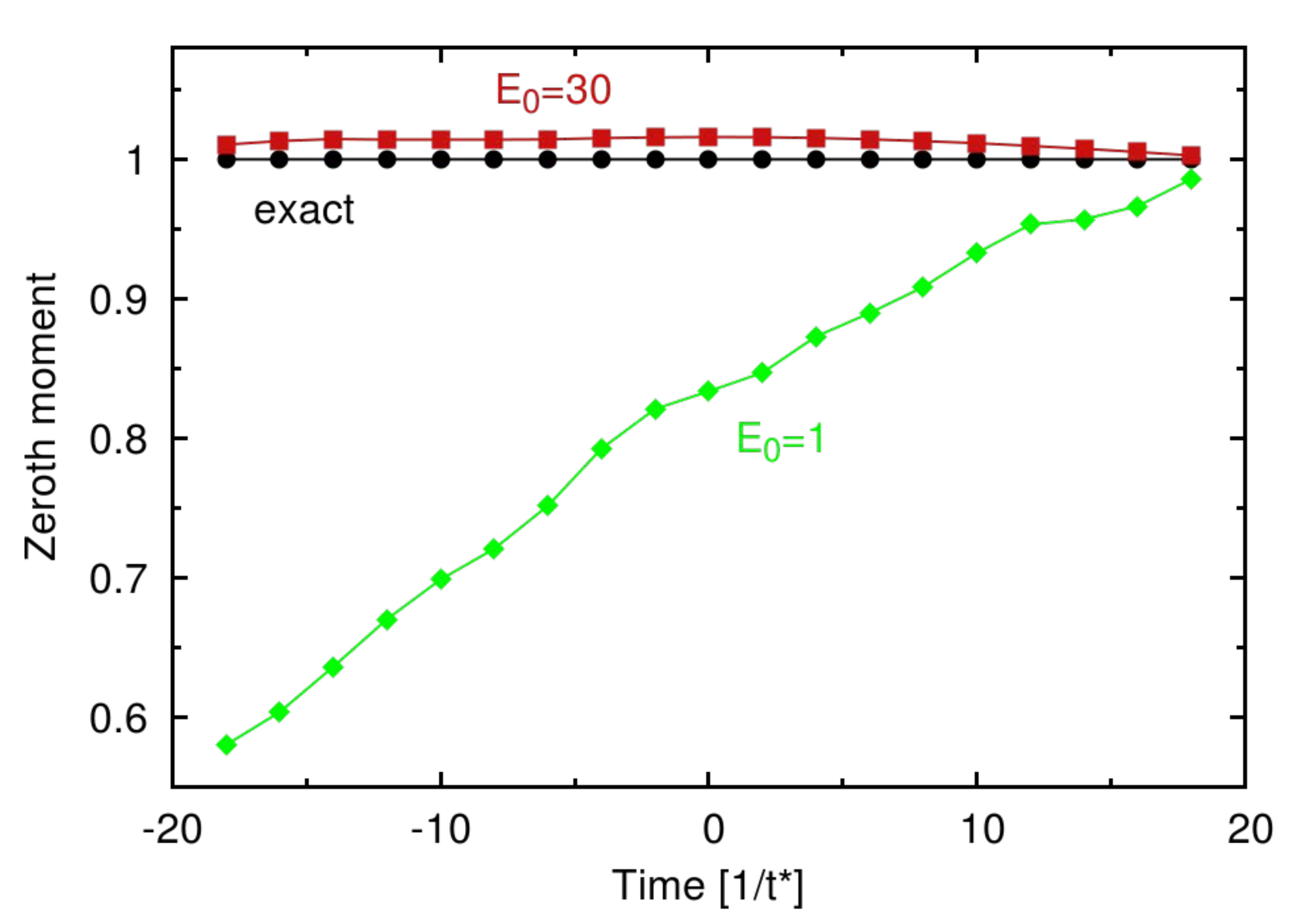}}
\caption{Comparison of the exact zeroth moment (black) to the zeroth moment for a large amplitude pulse ($E_0=30$, red) and a low amplitude pulse ($E_0=1$, green). Note how the sum rule is accurate within a few percent for the large amplitude case, but is quite poor for the smaller amplitude. [Figure reprinted from \cite{cdw_spie}, with permission]}
\label{fig: moments1}
\end{figure}

\begin{figure}[htb]
\centerline{
\includegraphics[width=0.6\textwidth]{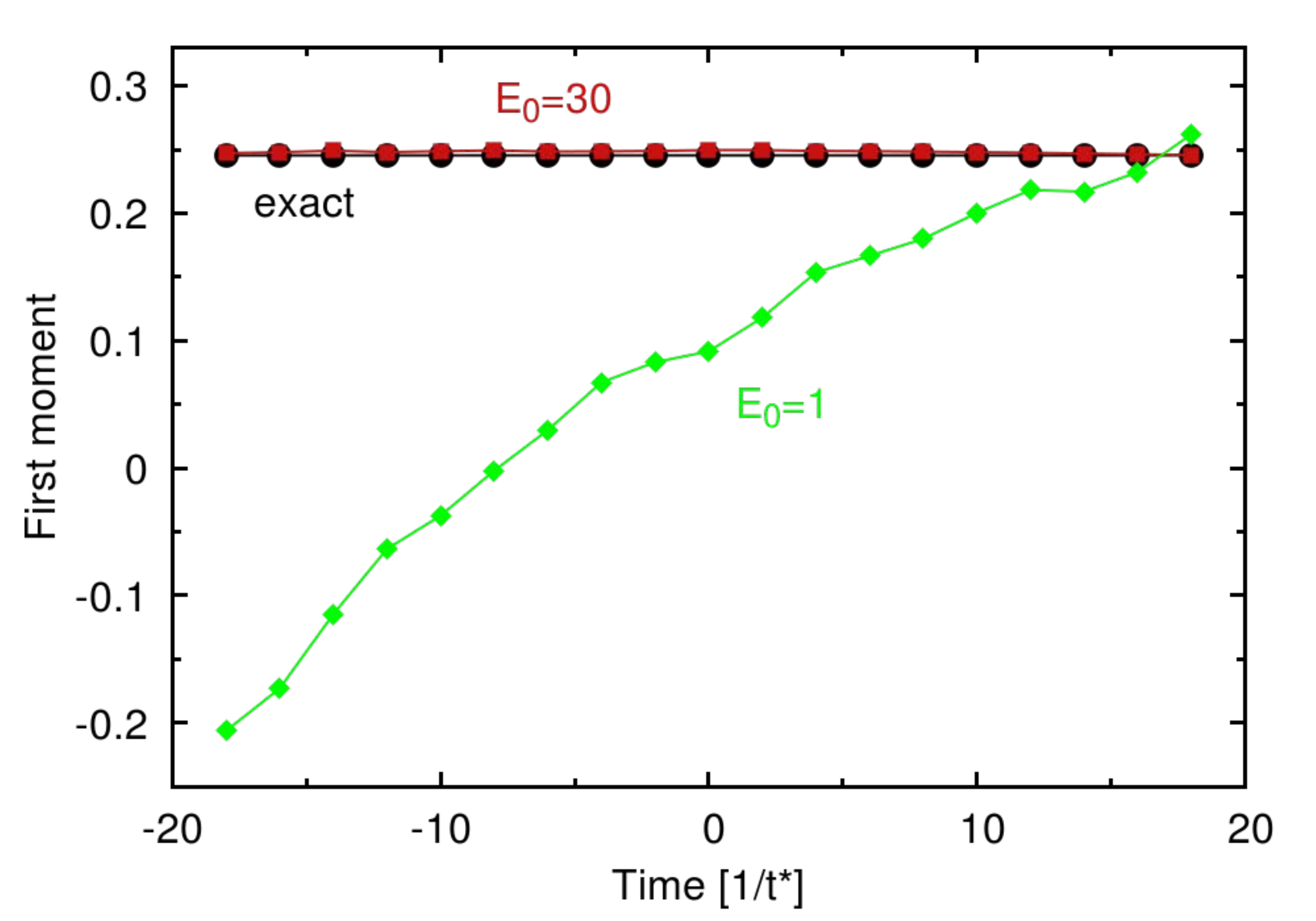}}
\caption{Comparison of the exact first moment (black) to the first moment for a large amplitude pulse ($E_0=30$, red) and a low amplitude pulse ($E_0=1$, green). Note how the sum rule is again accurate within a few percent for the large amplitude case, but is quite poor for the smaller amplitude.  [Figure reprinted from \cite{cdw_spie}, with permission]}
\label{fig: moments2}
\end{figure}

We end this section with a discussion of how to perform the numerical calculations.  In all cases, our goal is to determine the contour-ordered Green's function or the retarded and lesser Green's functions (which can be extracted from the contour-ordered one). The electric field is chosen to satisfy
\begin{equation}\label{efield}
\mathbf{E}(t)=\mathbf{E}_0\cos (\omega_p t)e^{-\frac{t^2}{\sigma_p^2}},
\end{equation}
where $E_0=|\mathbf{E}_0|$ is the magnitude of the field at time $t=0$.
For the simplified model we have already shown that the Green's function is directly found from the evolution operator, which decouples for each momentum. Furthermore, the retarded Green's function is determined solely by the relative time, so it typically does not require much computation to evaluate it. The lesser Green's function knows about the previous history of the system, so it requires longer runs in time to determine it, since we must start from a time in the distant past before the field is turned on. We have chosen, in this work, to evaluate the evolution operator via the Trotter formula. The only subtlety is how small of a time step do we take for each of the Trotter factors. This is then adjusted to ensure that the results have converged (best to extrapolate to $\Delta t=0$ and use sum rules to verify the convergence). One of the benefits of this approach is that we maintain unitarity explicitly for the evolution operator because each Trotter factor is determined analytically, and is manifestly unitary. An alternative way to solve this problem is to employ a conventional differential equation solver. The advantage of the differential equation  approach is that they can be made adaptive to help ensure appropriate accuracy, but they often suffer from loss of unitarity for long runs over large time intervals, and hence are often less reliable than the Trotter-based methods for these problems. We employed the Trotter approach for all results shown here.

\begin{figure}[htb]
\centerline{
\includegraphics[width=0.90\textwidth]{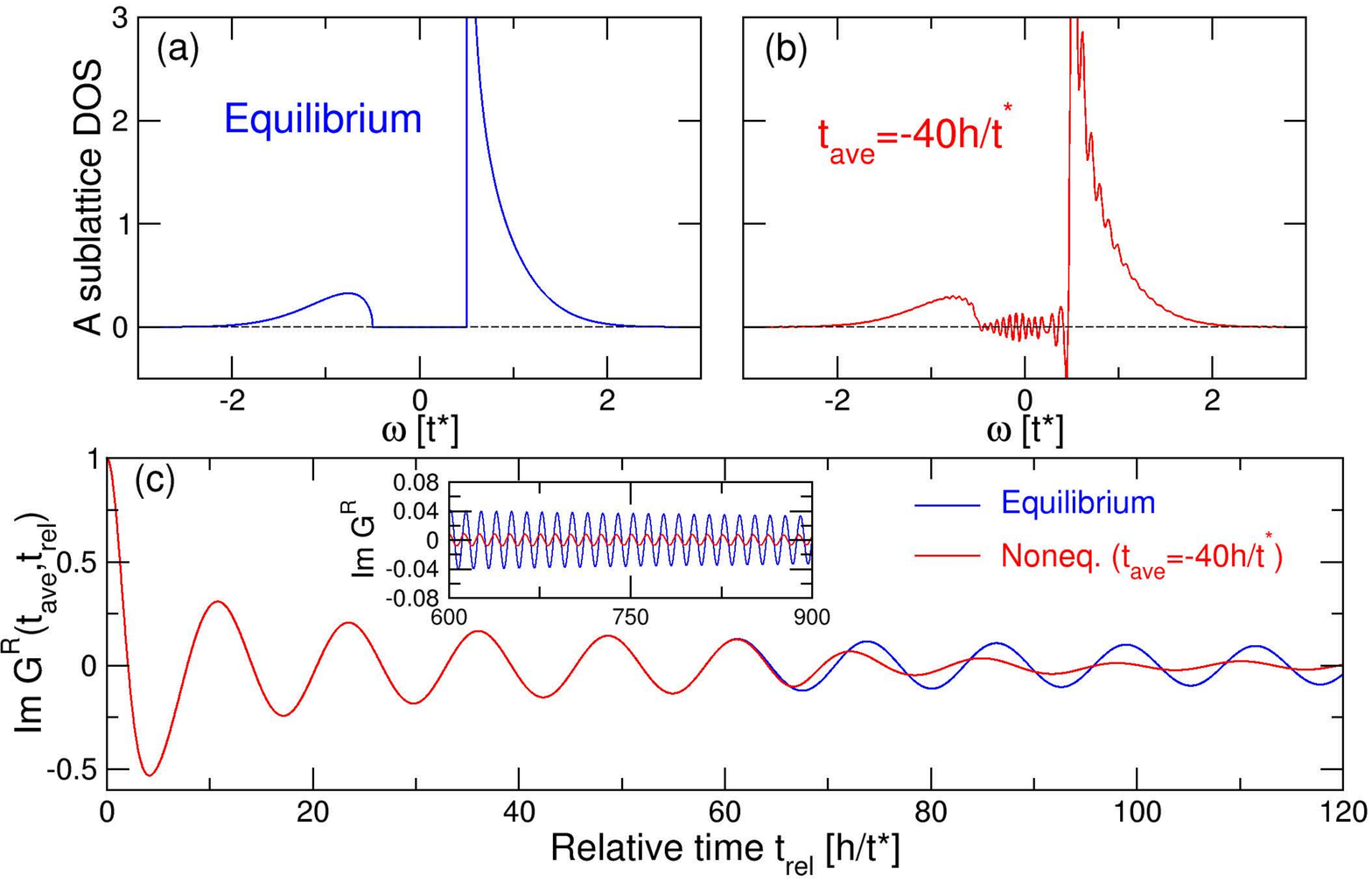}}
\caption{Density of states on the $A$ sublattice for (a) equilibrium and (b) nonequilibrium at an average time well before the pump pulse ($-40/t^*$). This is for the simplified model of the CDW with $U=1$, but also holds for the Falicov-Kimball model at $T=0$.
(c) Because the retarded Green's function has a long tail in the time domain, the density of states for a large range of average times is affected by the pump pulse. Inset, one can see the long tail of the Green's function which decays like $1/\sqrt{t}$ for the equilibrium Green's function, perhaps somewhat faster for the nonequilibrium Green's function. [Figure reprinted from \cite{cdw_pes}, with permission]}
\label{fig: cdw_dos}
\end{figure}

The Falicov-Kimball model calculations are more demanding, because they require the full nonequilibrium DMFT algorithm. We discretize the system, usually with $\Delta t= 0.066$, 0.05, and 0.033 and then quadratically extrapolate to zero $\Delta t$. As a check on the accuracy, we compute the zeroth and the first moment sum rules (the second moment accuracy is poorer during the initial part of the pulse). Empirically, we find that the equations converge more accurately when the amplitude of the pump pulse is large. This is illustrated in Figs.~\ref{fig: moments1} and \ref{fig: moments2}, where we plot the extrapolated moment sum rules for $E_0=1$ and $E_0=30$, and find that the large-amplitude case has acceptable errors, while the other has too large errors to be useful. It is surprising that this result is most accurate for the nonequilibrium regions; when the system is in equilibrium (left region of the figures), the errors are much larger. For this reason, we work exclusively with $E_0=30$ in this work for the Falicov-Kimball model results.

\section{Results}

We begin our discussion of the behavior of these systems by focusing on the density of states which will allow us to immediately discuss the phase diagram an quantum critical behavior. In both models, one can show that the $T=0$ DOS diverges as the inverse square root of the frequency at the band edges, which form the spectral band gap for the CDW. This is illustrated in Fig.~\ref{fig: cdw_dos}~(a) for the $A$ sublattice with the divergence at the upper band edge. When we Fourier transform this to relative time, the Fourier transform has a long tail which decays like $1/\sqrt{t}$. For an accuracy of 0.1\% in the DOS, one needs to run the relative time out to $1,000,000~1/t^*$ or more due to the long decay. This extreme nonlocality in time can cause misconceptions when one is working with the nonequilibrium system. In particular, if we fix the average time and Fourier transform with respect to the relative time, then once the relative time is large enough,
one of the two times in the $t$ and $t'$ basis will be earlier than the time when the field was applied and one will be later. Hence even for large negative average times, the DOS at that average time will be affected by the presence of the field, as depicted in Fig.~\ref{fig: cdw_dos}~(b) and (c). When Fourier transformed to frequency the DOS has significant oscillations that occur due to the slope discontinuity when the field is turned on (occurring near relative time of $60~1/t^*$ for this case). This analysis can only be done for the simplified model, because the Falicov-Kimball model cannot be calculated out to long enough times to see this behavior. This behavior is generic, however, for the transient DOS when the Green's functions in real time have long tails in equilibrium (in most cases these singularities, or sharp peaks disappear or are broadened in nonequilibrium, so the steady state DOS with a field present often have shorter tails in time, but the long-tails return for the quasiequilibrium states after the pump is turned on for a pump/probe experiment).

\begin{figure}[htb]
\centerline{
\includegraphics[width=0.75\textwidth]{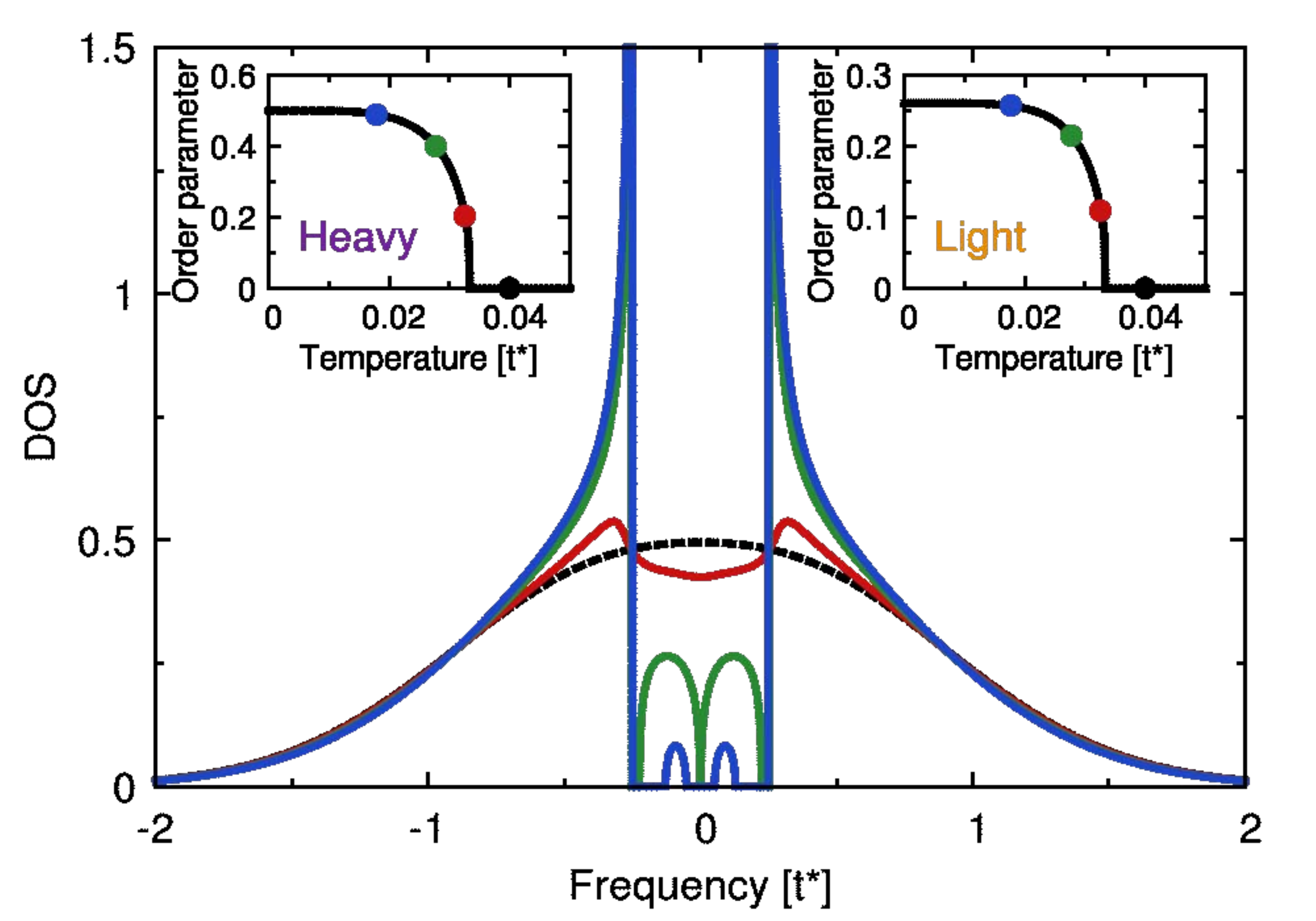}}
\caption{Average density of states for the Falicov-Kimball model with $U=0.5$, which corresponds to a strongly correlated metal. The curves correspond to different temperatures. Note how the singularity disappears at finite temperature and how the subgap states evolve. Inset is the order parameter for the
corresponding DOS, as indicated by the color. [Figure reprinted from \cite{cdw_prb}, with permission]}
\label{fig: cdw_dos_fk_metal}
\end{figure}

\begin{figure}[htb]
\centerline{
\includegraphics[width=0.75\textwidth]{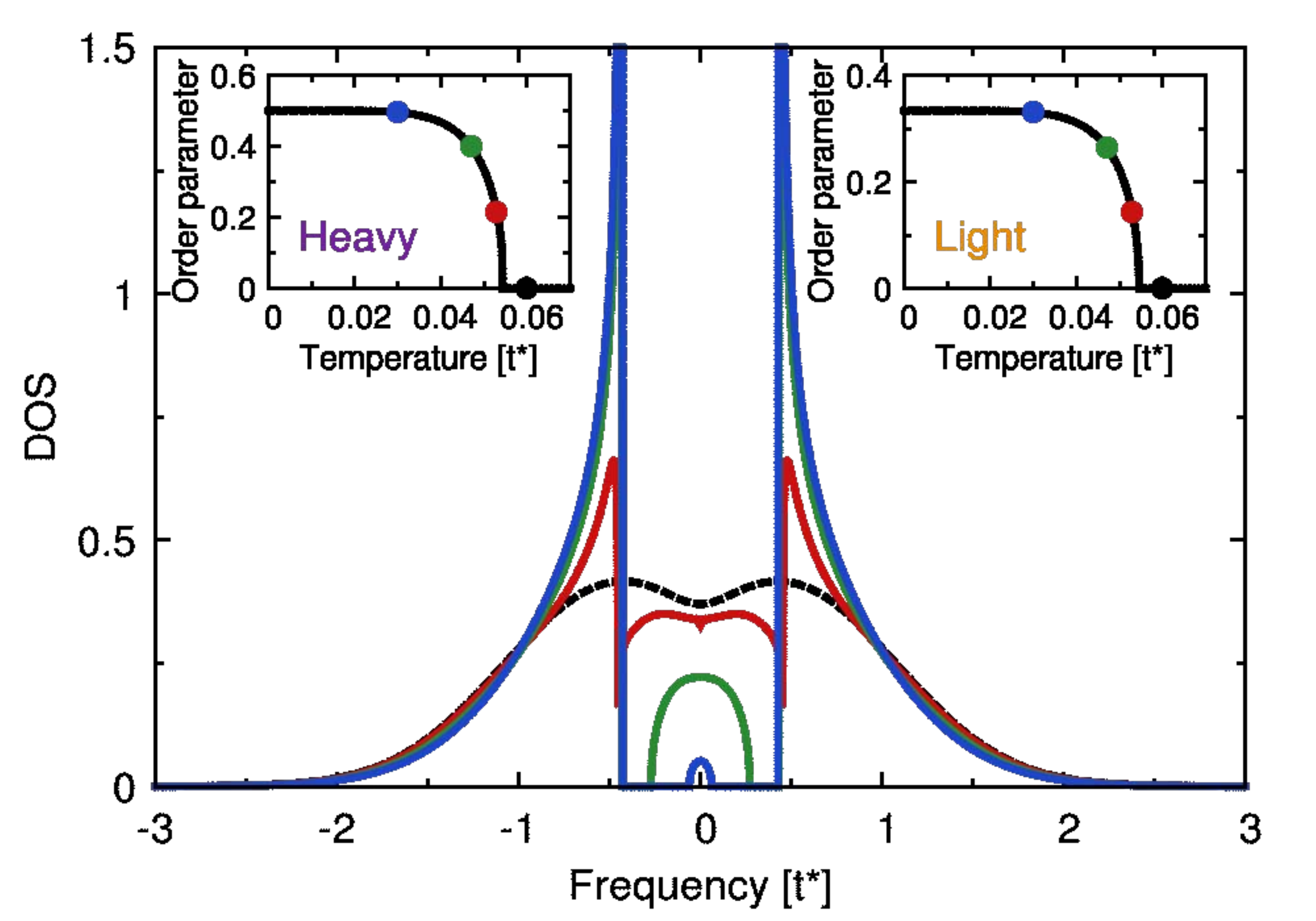}}
\caption{Average density of states for the Falicov-Kimball model with $U=0.86$, which corresponds to the quantum critical CDW. The curves correspond to different temperatures. Note how the singularity disappears at finite temperature and how the subgap states evolve. Inset is the order parameter for the
corresponding DOS, as indicated by the color. [Figure reprinted from \cite{cdw_prb}, with permission]}
\label{fig: cdw_dos_fk_crit}
\end{figure}

\begin{figure}[htb]
\centerline{
\includegraphics[width=0.75\textwidth]{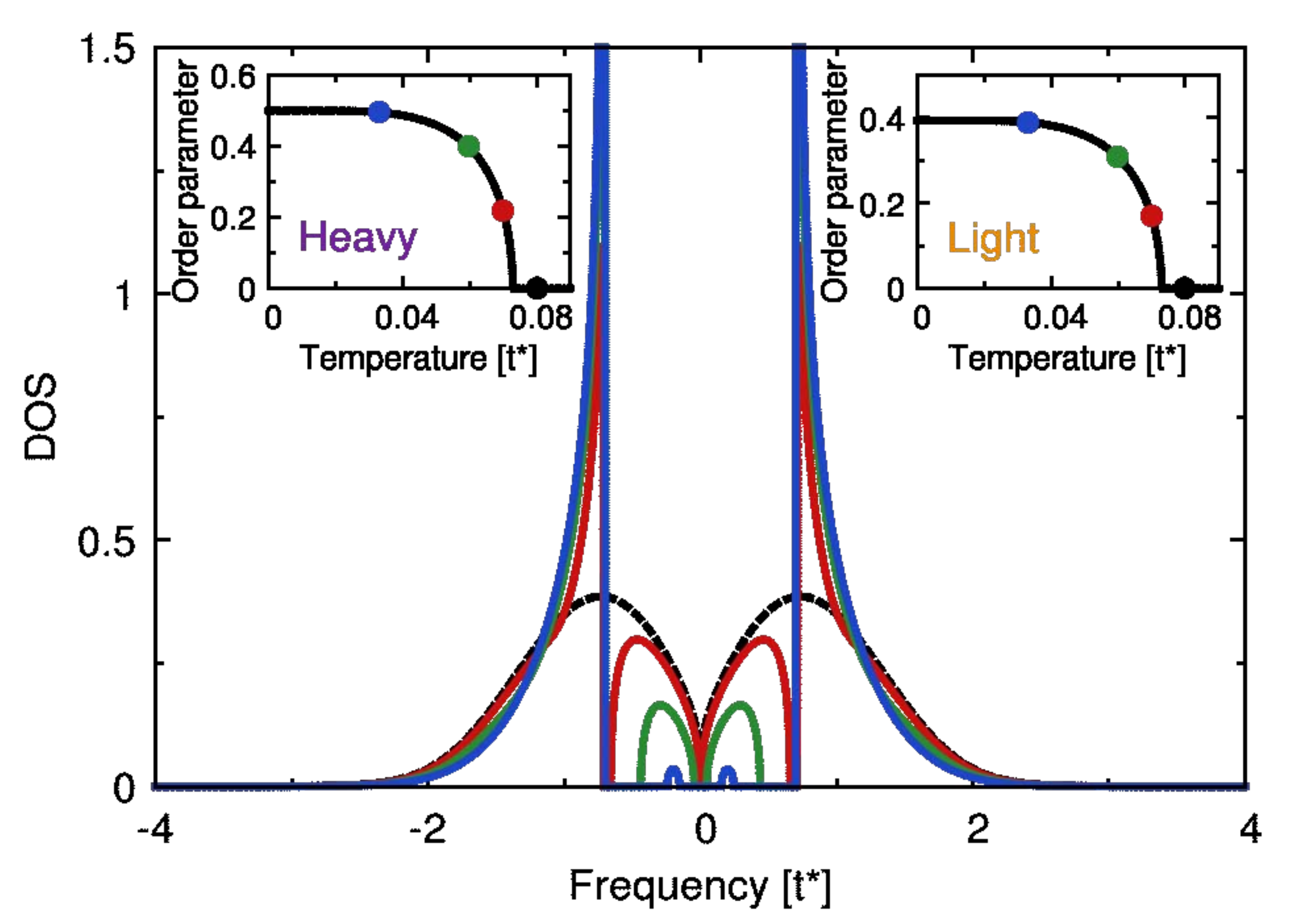}}
\caption{Average density of states for the Falicov-Kimball model with $U=1.4$, which corresponds to the critical point for the Mott insulator. The curves correspond to different temperatures. Note how the singularity disappears at finite temperature and how the subgap states evolve. Inset is the order parameter for the
corresponding DOS, as indicated by the color. [Figure reprinted from \cite{cdw_prb}, with permission]}
\label{fig: cdw_dos_fk_mott}
\end{figure}

We next focus on the behavior of the Falicov-Kimball model in equilibrium at finite-temperature, because the electron correlations bring on a distinctive behavior which is quite different from the standard Bardeen-Cooper-Schrieffer (BCS) paradigm~\cite{bcs}. In particular, if we define the spectral gap for the CDW to be the distance between the maxima that are the remnants of the divergence at $\omega=\pm U/2$ in the $T=0$ DOS, then we see that for most cases we will consider, the spectral gap remains fixed at $U$ all the way up until $T=T_c$ (or quite close to $T_c$). This differs completely from the BCS paradigm, where the spectral gap is tied directly to the order parameter (in that case the superconducting gap function), and the spectral gap shrinks as the order parameter shrinks until $T_c$ is reached where the spectral gap vanishes. Here, the phenomena is quite different. We instead have two subgap minibands that form once $T>0$, and they grow both in weight and in bandwidth until they close the subgap and then fill in the DOS from below as the spectral gap features diminish from above and both meet at $T=T_c$ to produce the normal state DOS. Figures~\ref{fig: cdw_dos_fk_metal},
\ref{fig: cdw_dos_fk_crit}, and \ref{fig: cdw_dos_fk_mott} plot the DOS in equilibrium for different temperatures for the metal, the quantum critical CDW and the critical Mott insulator phases. The different colors correspond to the different temperatures, with the dashed black line being the normal state. Inset are the order parameters for the light (right) and heavy (left) electrons. Note how the heavy electrons always have an order parameter that goes to 0.5 as $T\rightarrow 0$, but the light electrons order parameter is always less than that, although it increases as $U$ increases.

But the behavior is even more complex than this. The minibands initially start near the upper and lower band edges, but as $U$ increases, they migrate toward the center of the bandgap and they meet when $U=\sqrt{3}/2\approx 0.86602$. This is the underlying quantum critical point for the CDW, because at this point, the system has a transition from a metal to an insulator at $T\rightarrow 0$. The metallic phase opens as a fan for higher $T$ as the DOS becomes nonzero at the chemical potential and traces out the novel metallic CDW phase within the phase diagram. This phase is most stable for temperatures below but near $T_c$. As $U$ is increased further (up to $U=\sqrt{2}\approx 1.414$) we then have the Mott transition. In this case, the minibands form and grow in weight and broaden as $T$ increases, but they never broaden to completely fill the gap. Instead the band edges stop at the Mott insulator band edges, so the subgap region inside the Mott gap never fills with any states. Once one is above $T_c$, the DOS becomes temperature independent, and either has nonzero weight at the chemical potential for the metal or has no DOS at the chemical potential for the Mott insulator.

\begin{figure}[htb]
\centerline{
\includegraphics[width=0.65\textwidth]{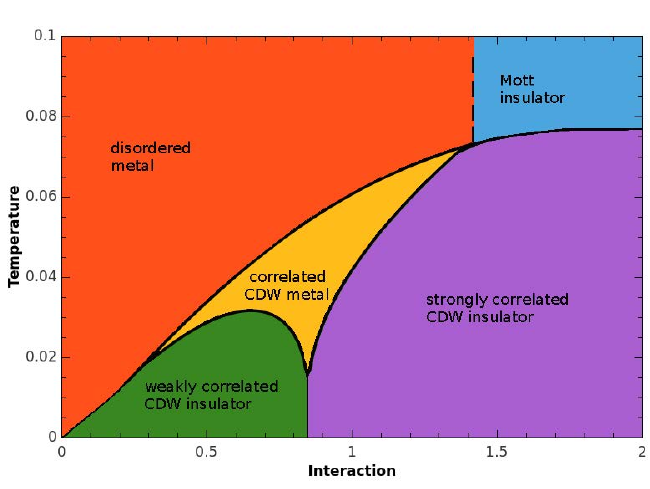}}
\caption{Phase diagram for the Falicov-Kimball model when both electron densities are 0.5. In the normal state, there is a transition from a metal to a Mott insulator at $U=1.4$, characterized by the opening of a gap in the single-particle density of states, which is the same on each sublattice. In the CDW ordered phase, where the densities of the particles are different on the two sublattices, there are three phases: (1) a weakly correlated CDW phase, which is continuously connected to $U=0$;
(2) a strongly correlated CDW metal, which is present only at nonzero temperature, and emerges from the
quantum critical point of the model at $U=0.86$; and (3) a strongly correlated CDW insulator, which is continuously connected to the Mott insulator within the CDW phase. The quantum critical point is nonstandard, because the order parameter for the CDW order varies continuously through the transition.}
\label{fig: cdw_phase}
\end{figure}

The phase diagram showing all of these phases and their region of stability is plotted in Fig.~\ref{fig: cdw_phase} for the spinless Falicov-Kimball model on a hypercubic lattice. The five different phases are indicated by the different colors and the boundary lines. Note how the quantum critical CDW region is extremely narrow for temperatures below 0.02~$t^*$. It becomes quite difficult to determine the precise phase boundary in this region because it is hard to calculate the DOS accurately there. The phase boundaries are more rigorously determined and the existence of the quantum critical point is explicitly proved on an infinite-coordination Bethe lattice~\cite{lemanski} (as opposed to the hypercubic lattice results shown here). Nevertheless, we work with the hypercubic lattice here, where the phase boundary is more challenging to explicitly determine.

Note that this quantum critical point is rather unique. An ordinary quantum critical point will occur at the terminus of a phase diagram where an order parameter is suppressed to zero. At that point, there is a zero-temperature phase transition. In many quantum critical points, the original quantum critical region is hidden by a superconducting dome. The case here is different for a number of reasons. First, the spatial order parameter does not disappear at the quantum critical point, instead, it varies continuously at the critical point. Instead, the DOS has a change of character as $T\rightarrow 0$ versus $T=0$, the former being metallic and the latter
being insulating. Nevertheless, there remains a strange metal fan above the quantum critical point, similar to
more conventional quantum critical points.

\begin{figure}[htb]
\centerline{
\includegraphics[width=0.55\textwidth]{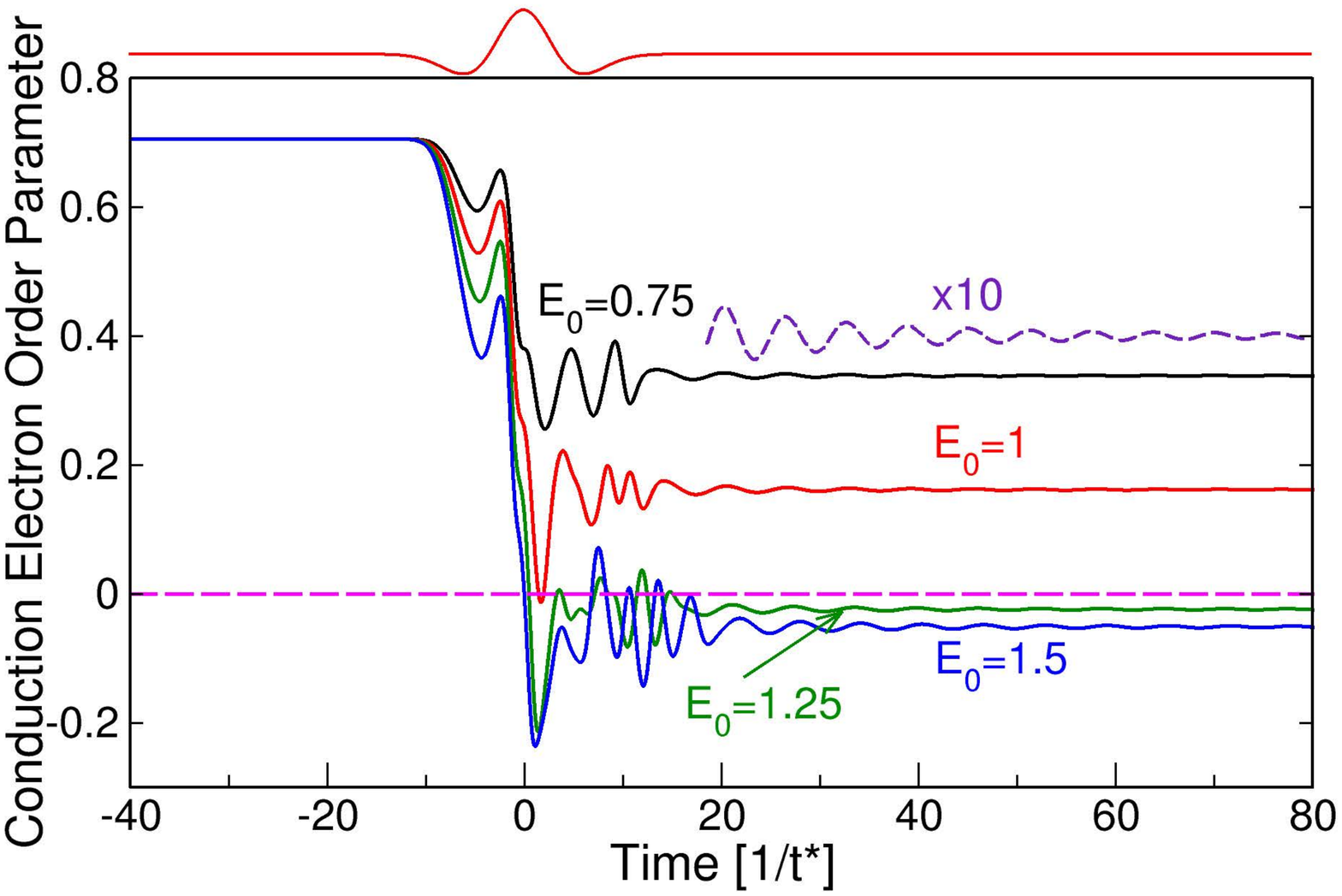}}
\caption{
Conduction electron order
parameter for the simplified model of a CDW with $U=1$ as a function of time for different
pulse amplitudes (zero is indicated by the magenta dashed
line). The purple dashed line is a zoom in of the black curve for long times. The field is shown at the top 
in red.
[Adapted from \cite{cdw_pes}, with permission]}
\label{fig: cdw_non_order}
\end{figure}

Now we turn to describing the conduction electron order parameter and how it varies with the
pump pulse. As electrons are driven by the field, they will flow through the material moving from the $A$ to the $B$ sublattice and being excited up to the higher energy band (or de-excited to the lower energy band). As a result, we anticipate that the conduction-electron order parameter will transiently change as a function of time. This is indicated for the simplified CDW model in Fig.~\ref{fig: cdw_non_order}. The cases with a pulse amplitude of 0.75 and 1 both oscillate and are pushed downwards, but settle into values that are positive for the
long-time limit, while, when the amplitude is increased to 1.25 and 1.5, the order parameter actually changes
sign! For larger amplitudes we anticipate that the order parameter will continue to oscillate and the sign of the long-time limit may be difficult to determine without running through the whole calculation.

\begin{figure}[htb]
\centerline{
\includegraphics[width=0.495\textwidth]{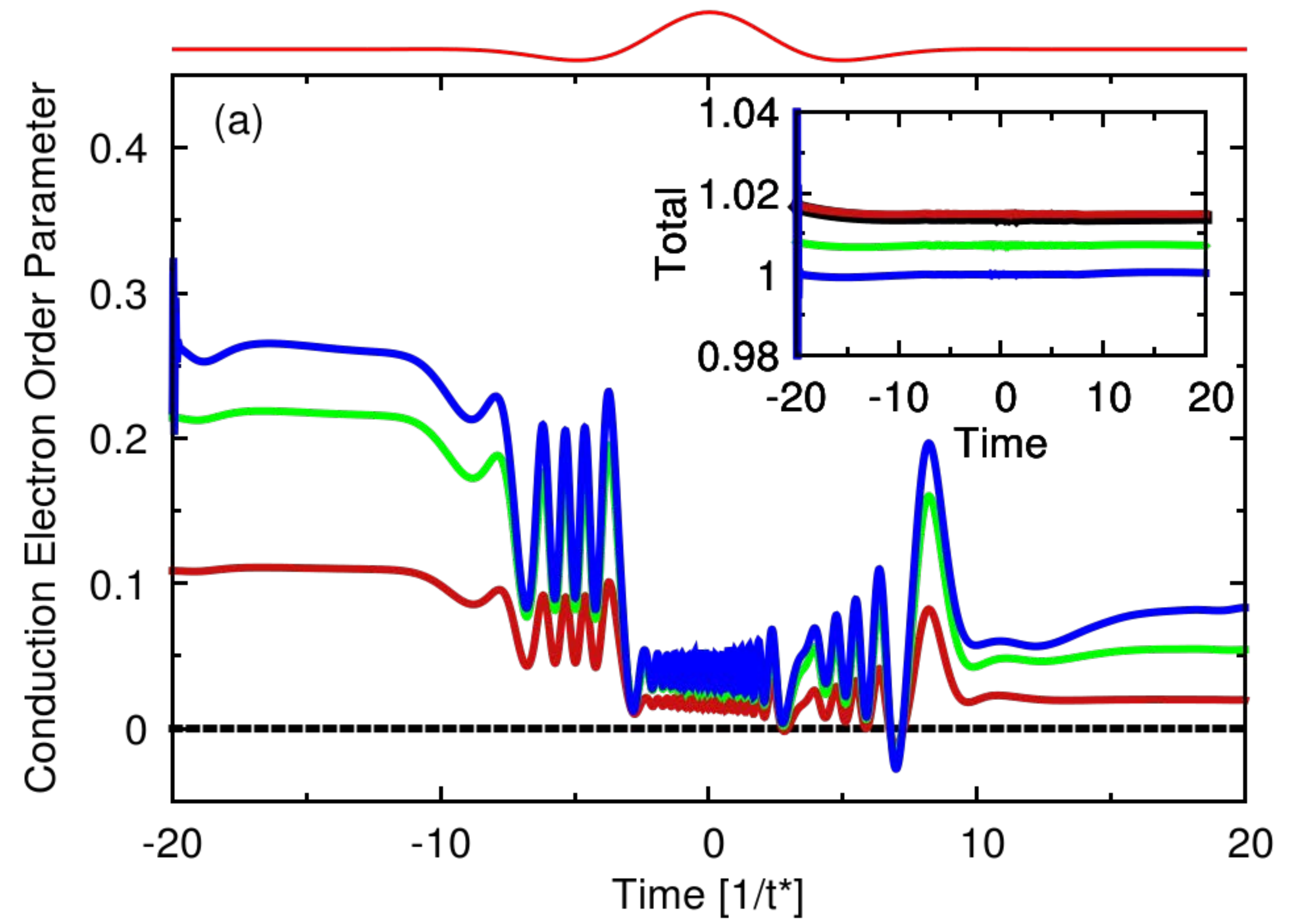}
\includegraphics[width=0.495\textwidth]{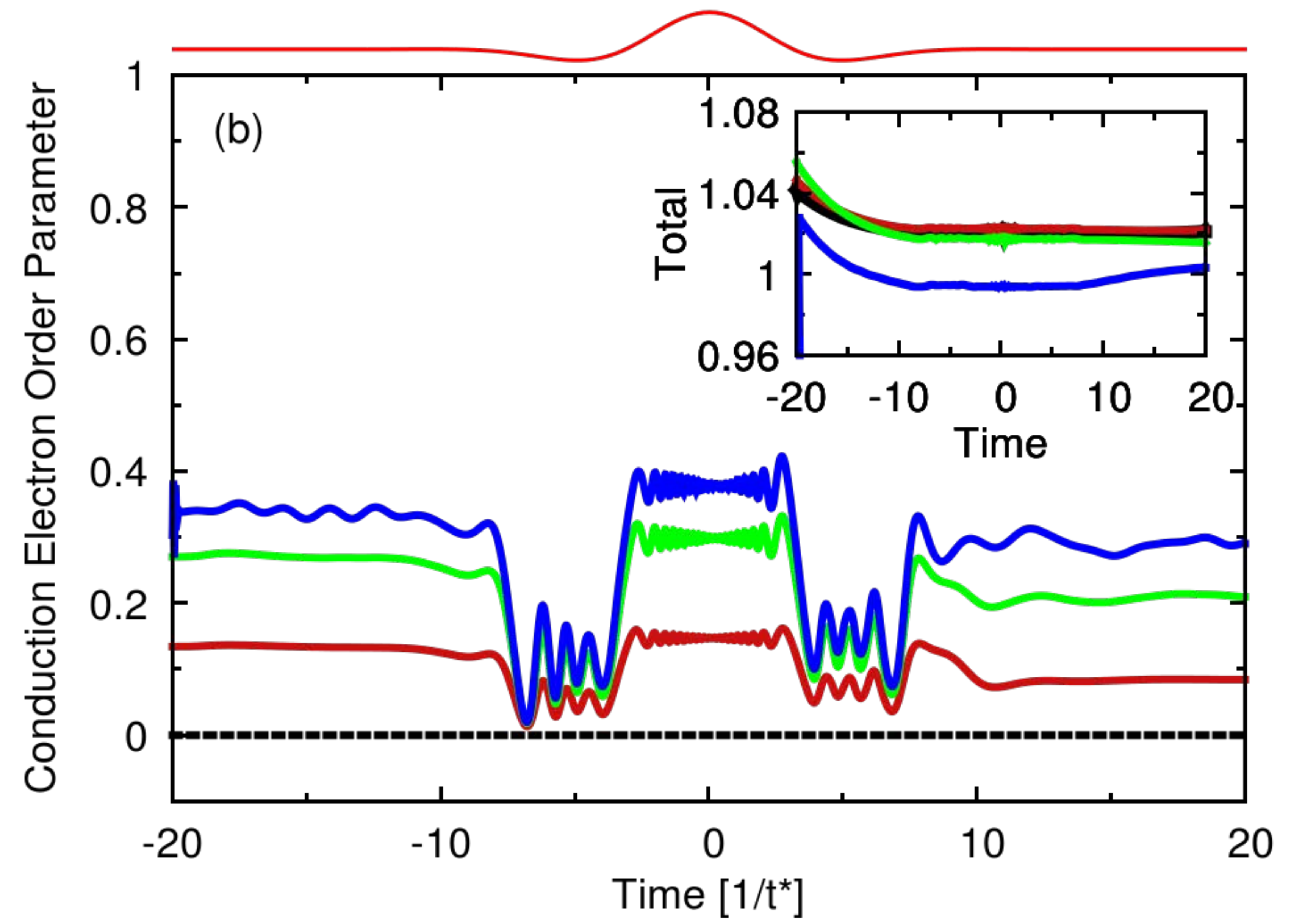}}
\centerline{
\includegraphics[width=0.495\textwidth]{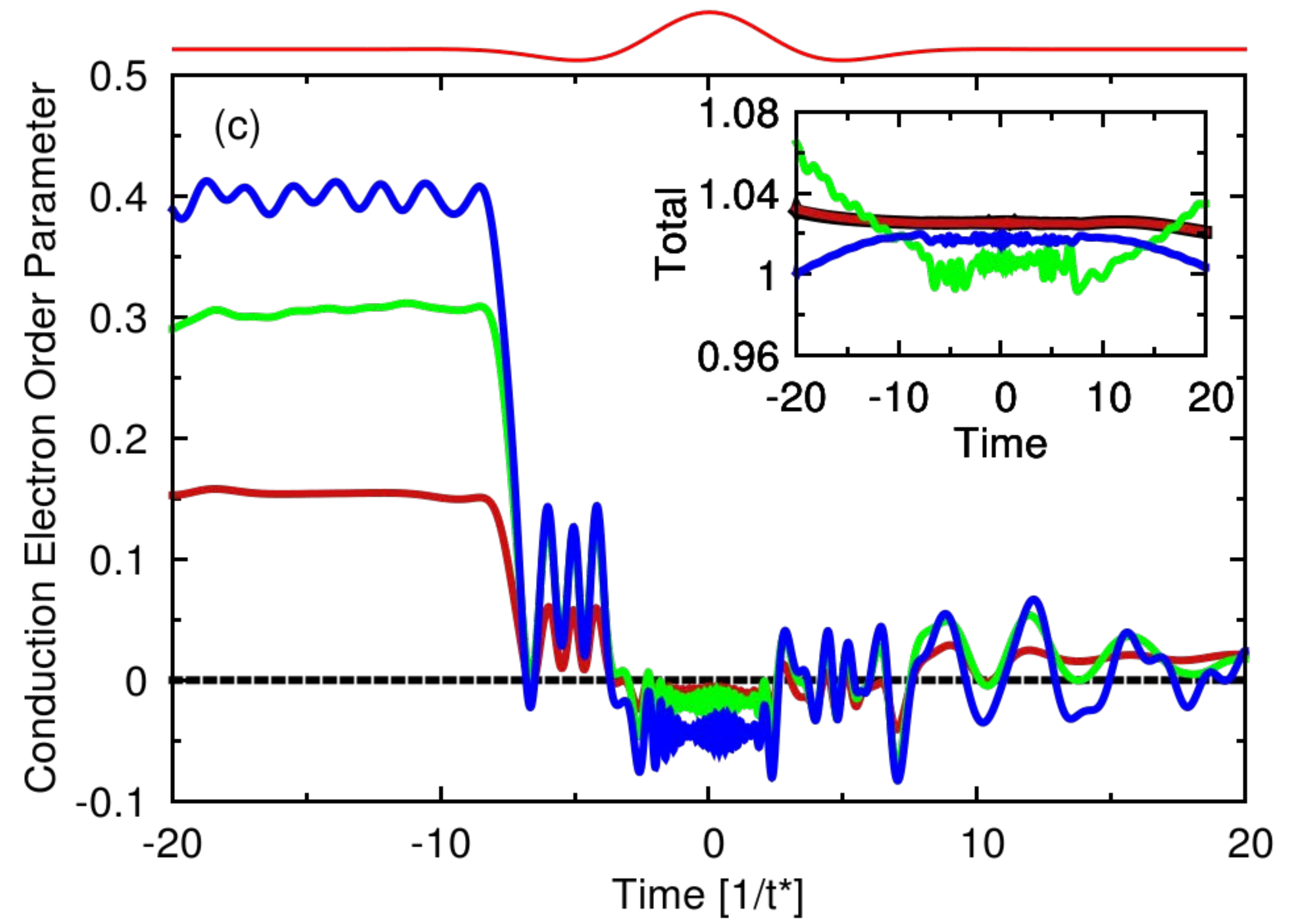}}
\caption{
Conduction electron order
parameter for the CDW as a function of time for different
initial temperatures in the Falicov-Kimball model for the (a) metal, (b) quantum critical point, and (c) critical Mott insulator.
The inset shows the total filling. Errors are on the order of a few percent due to
difficulty in scaling the numerics; for example, the order parameter should be constant prior to the field
being turned on, but we see some small wiggles, illustrating the accuracy of the calculation.
The field is shown above in red.
}
\label{fig: cdw_fk_order}
\end{figure}

In Fig.~\ref{fig: cdw_fk_order}, we show a similar figure, but this time for a large amplitude pulse ($E_0=30$) in the Falicov-Kimball model with different initial temperatures. The simplified CDW model will be closest to the 
low-temperature results. Initially, the system starts in equilibrium, and so the order parameter should be a constant, but the numerical results show some small dependence on time in this region (on the order of a few percent) which is an indication of the accuracy of the data after the extrapolations have been done. Inset, we show the total electron filling, which should be precisely 1 throughout the simulation. Its variation is another indication of the accuracy of these calculations, which are pushing the state-of-the-art to its limits.

The metallic case is in panel (a) and it shows rather expected behavior. It starts off fairly flat, is reduced as the field is turned on, with large oscillations, and then settles into a rather flat result in the long-time limit, which is reduced from the original equilibrium value, but remains positive.  Panel (b) of the quantum critical CDW is much more interesting. It shows a nearly symmetric curve, where the order parameter is reduced, but in the end, the final value is quite similar to the initial equilibrium value. We will see that this occurs because it is difficult to excite the quantum critical CDW. Finally, panel (c) shows the critical Mott insulator where in all cases, regardless of the starting point, the order parameter is being suppressed almost to zero. It is surprising that this suppression is more prominent for the Mott insulator than it is for the metal, and there is no simple explanation for why this would be so.

\begin{figure}[htb]
\centerline{
\includegraphics[width=0.45\textwidth]{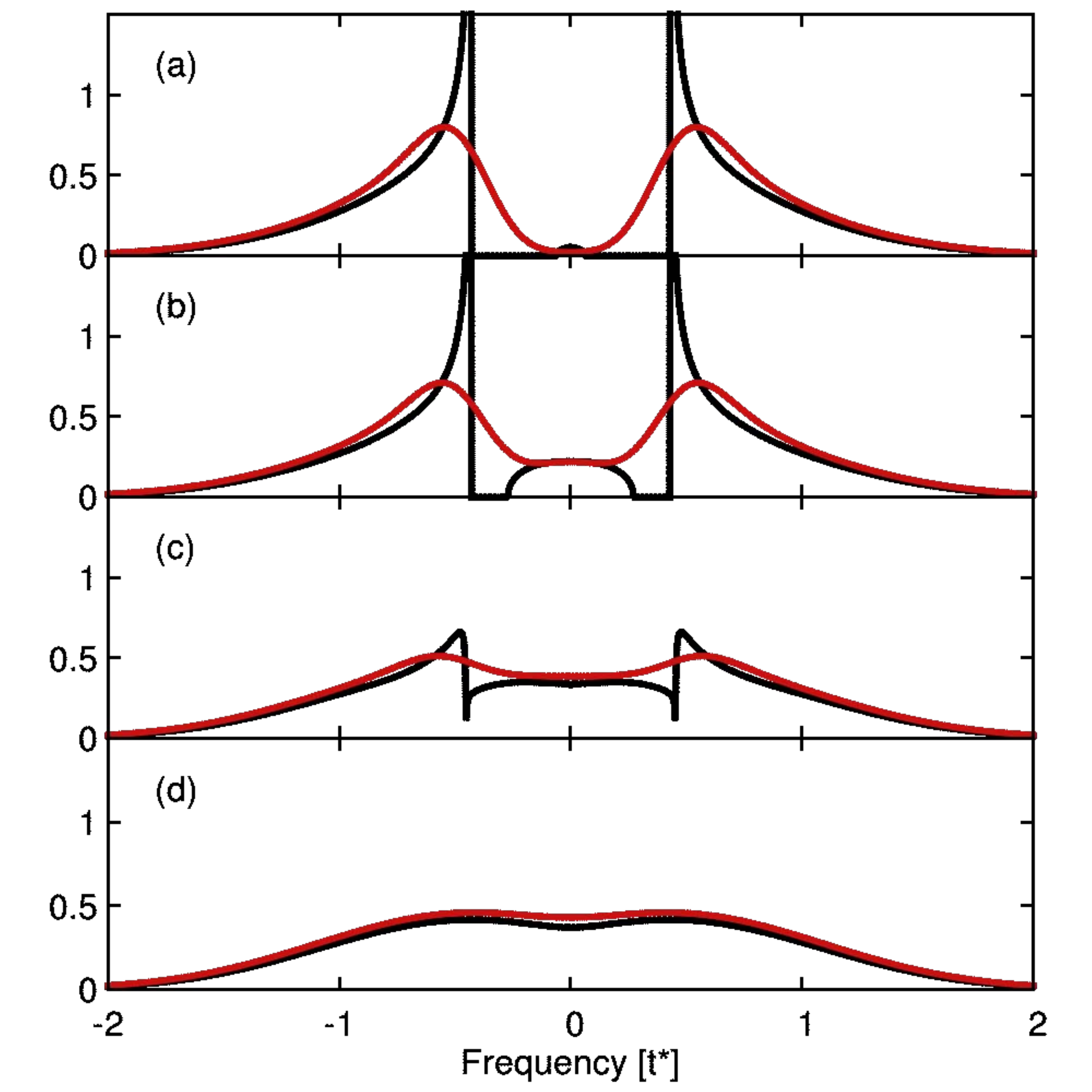}
\includegraphics[width=0.45\textwidth]{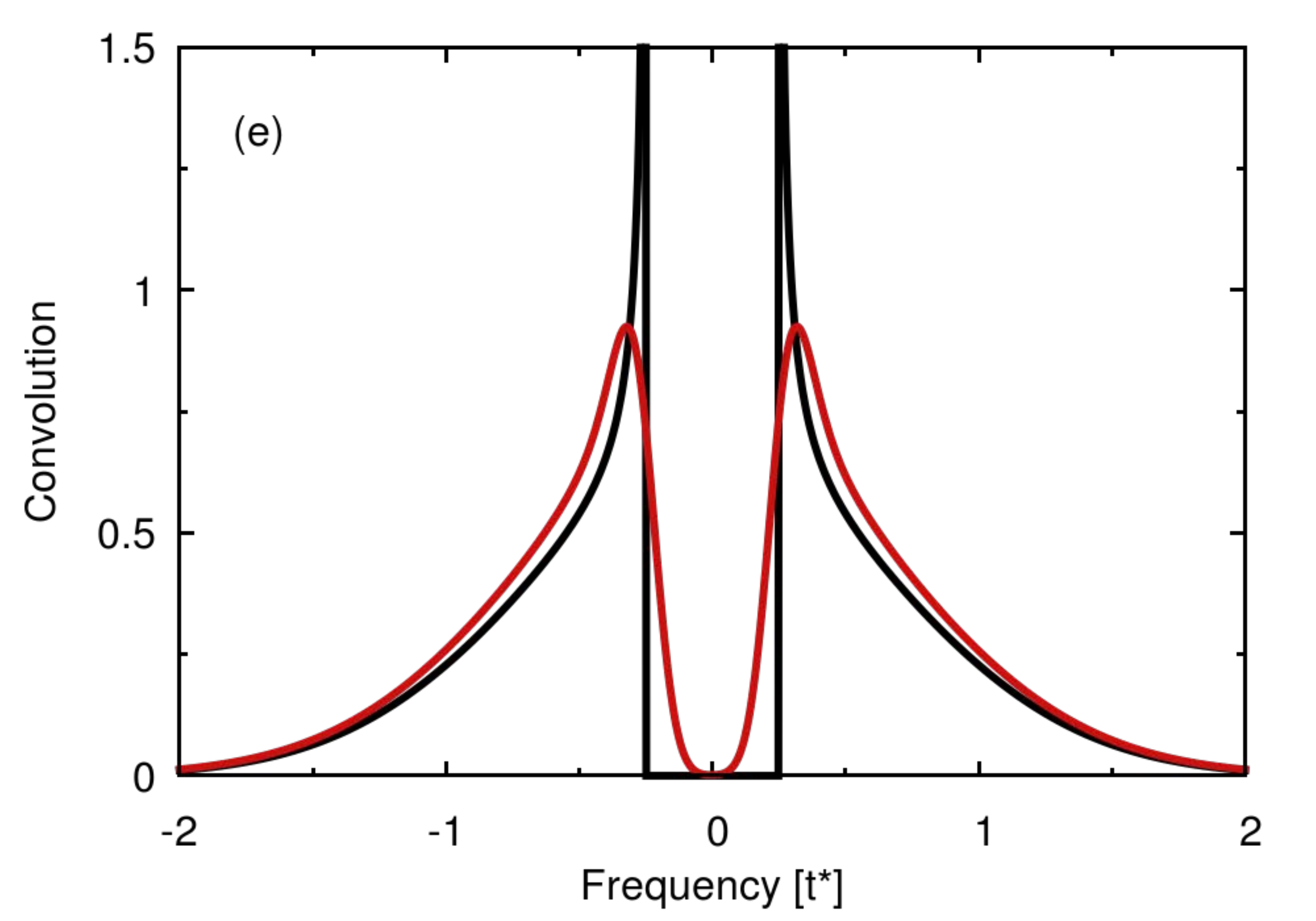}}
\caption{
(Panels a-d) Convolution of the equilibrium density of states (red) and the density of states (black) for the quantum critical CDW in the Falicov-Kimball model
with a Gaussian of width $\sigma_b=7t^*$ and different temperatures. The convolution washes out a number of the features in the 
density of states, which will continue as the system is pumped. Panel (e) shows the same for the simplified model with $U=0.5t^*$ and $\sigma_b=14t^*$.
[Panels (a-d) reprinted from \cite{cdw_prb}, with permission]}
\label{fig: dos_conv}
\end{figure}

The formula for the time-resolved photoemission spectroscopy in Eq.~(\ref{eq: pes}) involves a double-time Fourier transform weighted by the probe envelope given in Eq.~(\ref{eq: probe}). The form of the photoemission integral is that of a convolution. When the probe width $\sigma_b$ is narrow, one has
high time resolution but poor spectral resolution and {\it vice versa} when the probe width is broad. To understand the effect this has on the features of the density of states, we show the convolution of the 
probe function used for the quantum critical CDW state of the Falicov-Kimball model, where $\sigma_b=7t^*$ with the local DOS for different
temperatures in Fig.~\ref{fig: dos_conv}~(a--d). One immediately sees that the gap region is smoothed out, and the details of the subgap states are lost. Furthermore, as one approaches the normal state, the convolution becomes quite close to the original signal because it does not have any sharp features. When the interaction is decreased to $U=0.5$ and the probe width is increased to $\sigma_b=14t^*$, as shown for the $T=0$ data on the right, we see that the smoothing out of the features is much less and the gap is nearly fully formed. The divergent peaks however remain smoothed out, because they require long tails in time before they are fully developed. It is important to keep in mind the smoothed out features due to the convolution when we look at the time-resolved PES next, because those data will also have the smoothed out behavior.

\begin{figure}[htb]
\centerline{
\includegraphics[width=0.49\textwidth]{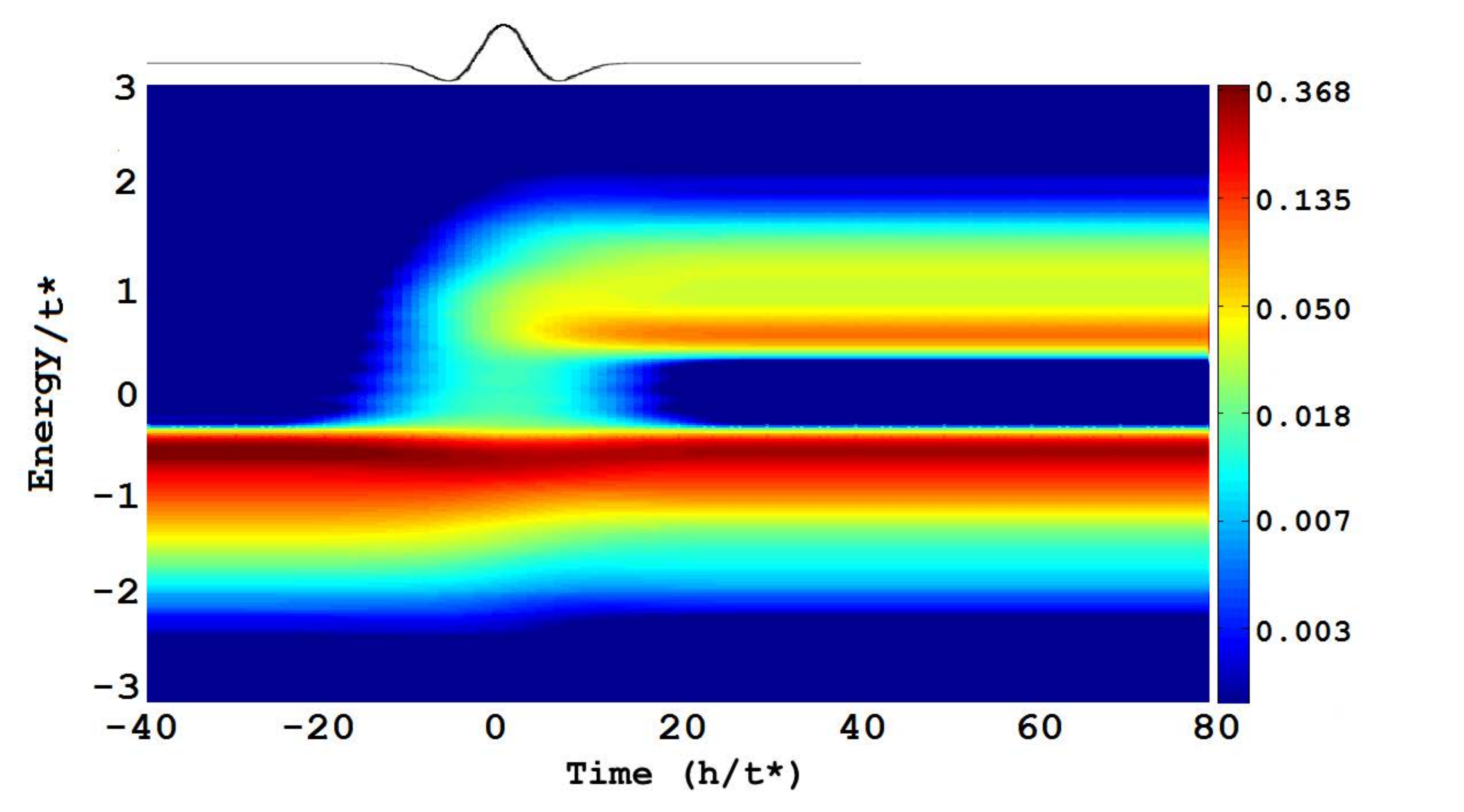}}
\caption{
False color plot of the calculated time-resolved PES at $T = 0$
with $E_0 = 0.5$ and averaged over the A and B sublattices,
for the simplified CDW model. The electric field profile is shown above the
plot.
[Reprinted from \cite{cdw_pes}, with permission]}
\label{fig: pes_non}
\end{figure}

The results for the transient time-resolved PES for the simplified model with $E_0=0.5$ is shown in Fig.~\ref{fig: pes_non}. The false-color plot has a logarithmic scale for the PES to emphasize the data at small values. One can immediately see that the gap closes (light blue region near $t=0$) when the field is present and then re-opens. A substantial number of electrons are excited by the pulse, and there is a small band narrowing, but the spectral gap remains at about the same value for all times (perhaps it shrinks by a percent or two). Since the noninteracting system requires a field to be present for both excitation and de-excitation, the number of electrons excited into the upper band does not change once the pump pulse is over. We will use the same
false-color scale for all of the PES shown in this work.

\begin{figure}[htb]
\centerline{
\includegraphics[width=0.49\textwidth]{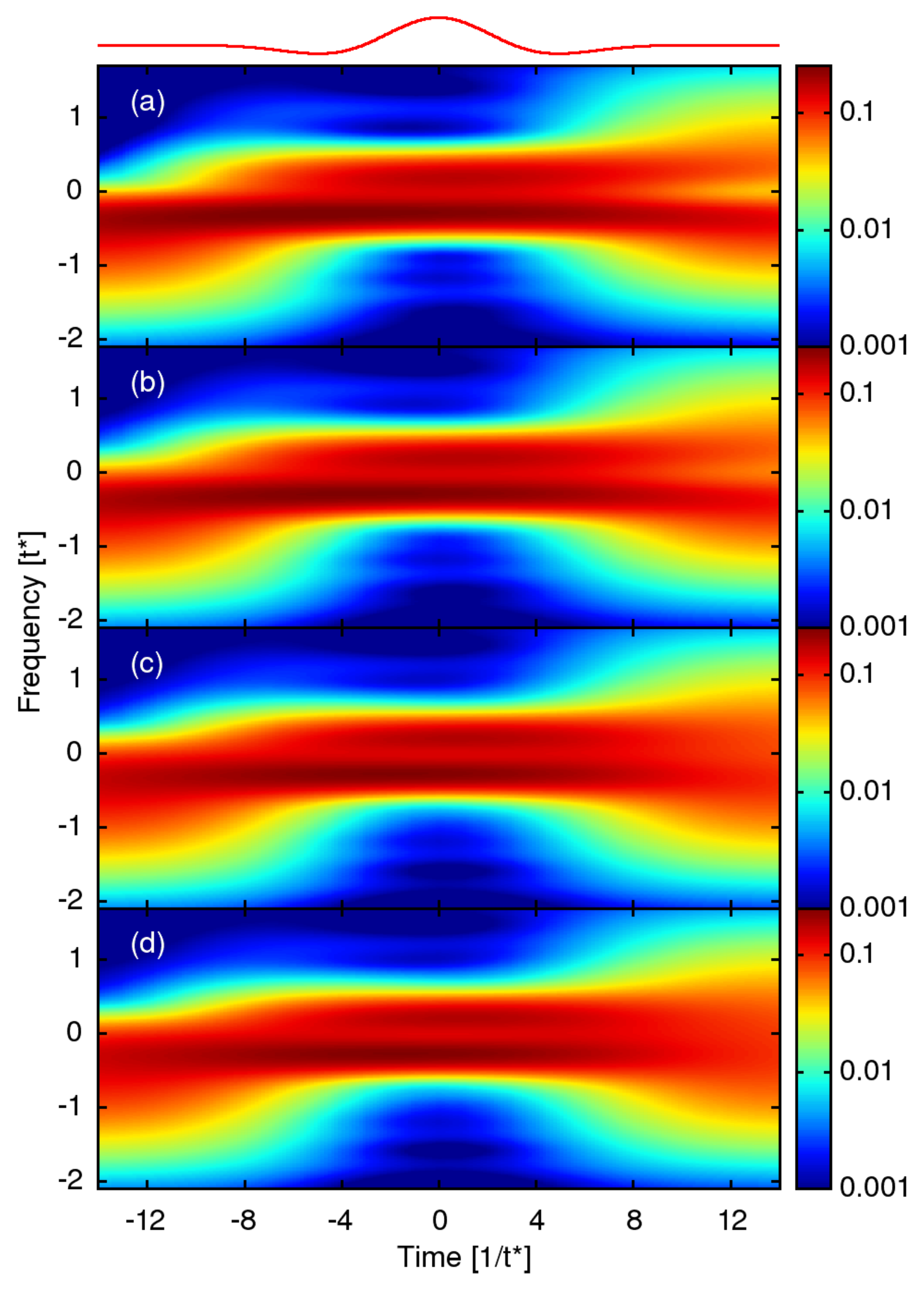}
\includegraphics[width=0.49\textwidth]{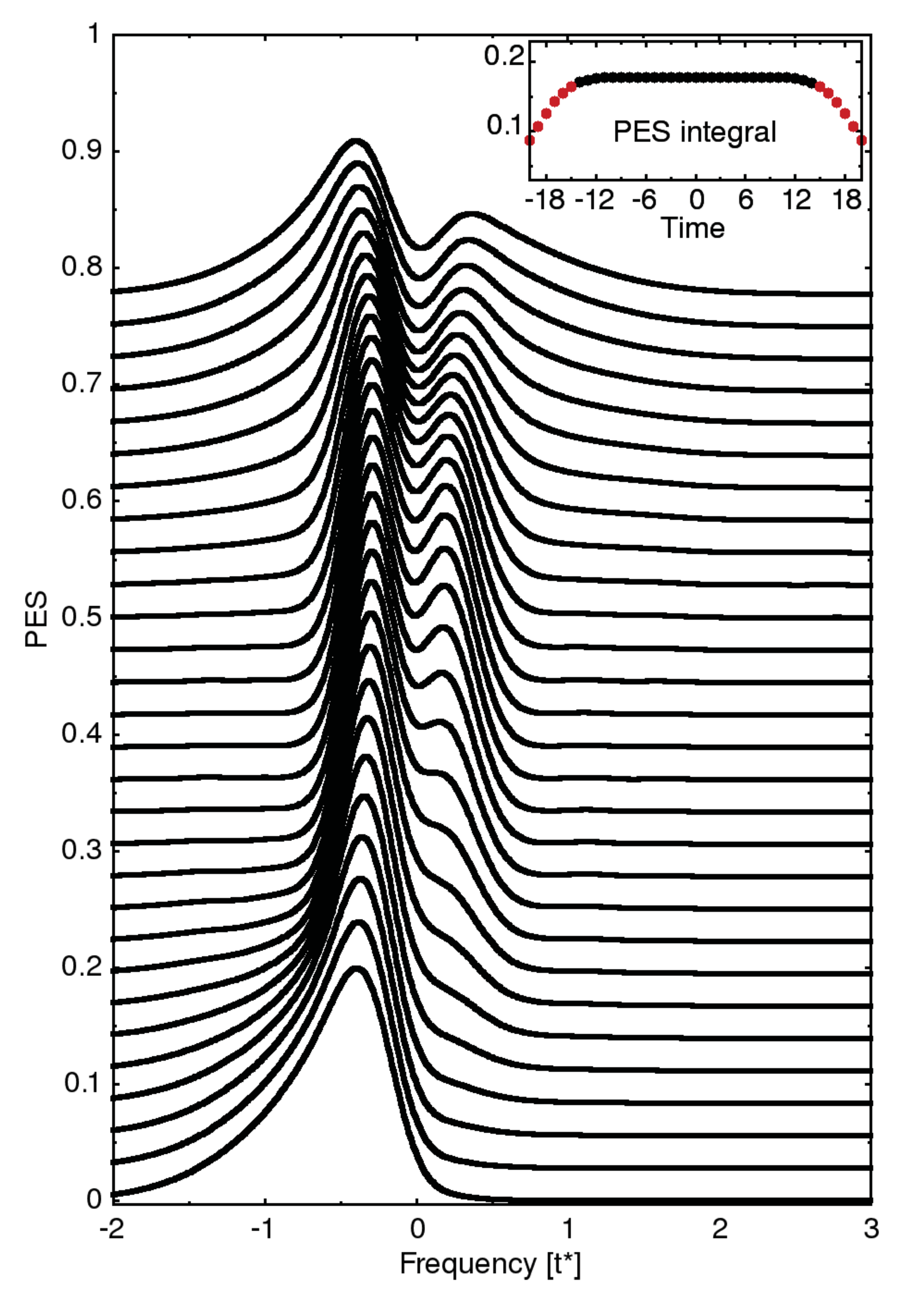}}
\caption{
(a--d, left) False-color plot of the calculated time-resolved PES for various $T$
with $E_0 = 30$, $U=0.5$ and averaged over the A and B sublattices,
plotted in false color. This system is a strongly correlated metal. The electric field is shown above the
plot.  The different panels are for the different temperatures in Fig.~\ref{fig: cdw_dos_fk_metal} with (a) the lowest and (d) the highest temperature.  (right) Waterfall image of vertical cuts through the TR-PES data plotted for different delay times and
offset for clarity. Inset is the total integrated spectral weight. Only the data that conserves the spectral weight is shown. This data corresponds to the case of panel (a) to the left.
[Reprinted and adapted from \cite{cdw_prb}, with permission]}
\label{fig: pes_fk_metal}
\end{figure}

We now move on to the Falicov-Kimball model, where we are required to run the calculations with a much 
larger amplitude of the field, otherwise the extrapolated results have not yet converged.  The data given here
all correspond to $E_0=30$. In Fig.~\ref{fig: pes_fk_metal}, we show a series of false color images on the left and a waterfall plot for the lowest temperature on the right. There are a few features to emphasize here which are different from what we saw in the simplified CDW case. Some of these may be arising from the fact that the field amplitude is so large now. First, we see a significant narrowing of the band when the field is on (notice the 
large blue region near $t=0$). Because the total spectral weight is conserved, this narrowing comes with a sharpening of the peaks in the PES as well, which is a bit harder to see in the false color image, but may be clearer in the waterfall. One can also clearly see that the spectral gap is shrinking, as the lower gap edge is being pushed toward 0 in the waterfall. We inset the total spectral weight as a function of the probe delay, to show that the data we use has conserved spectral weight, while for too long or too short delays, our data becomes poor, and spectral weight is lost. We do not show any data for the red points here. The severe band narrowing is quite surprising, as it is coming from a dressing of the electronic states by the pump pulse. The surprise is that the bandwidth is narrowed almost by a factor of two, which is a significant effect.

\begin{figure}[htb]
\centerline{
\includegraphics[width=0.49\textwidth]{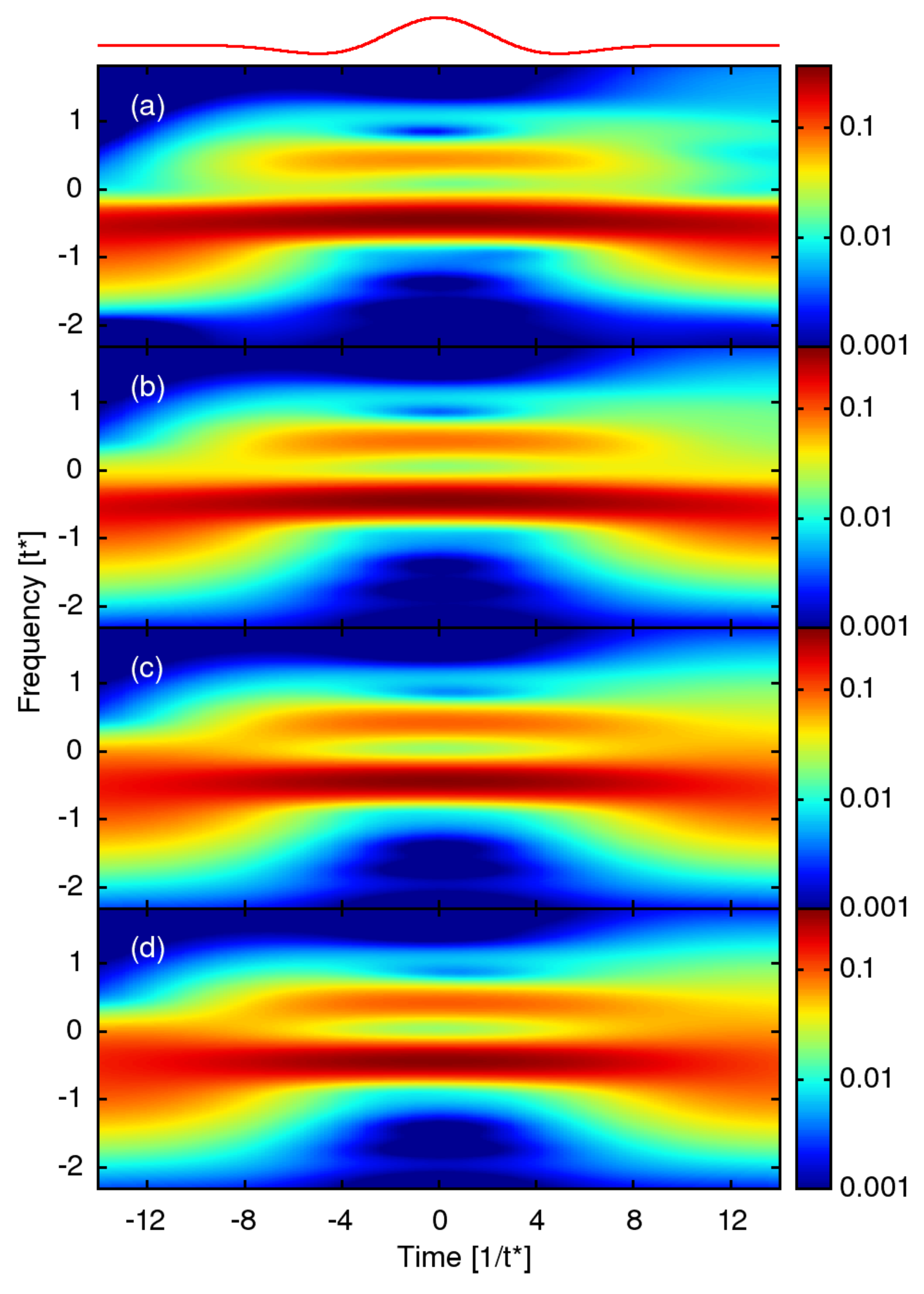}
\includegraphics[width=0.49\textwidth]{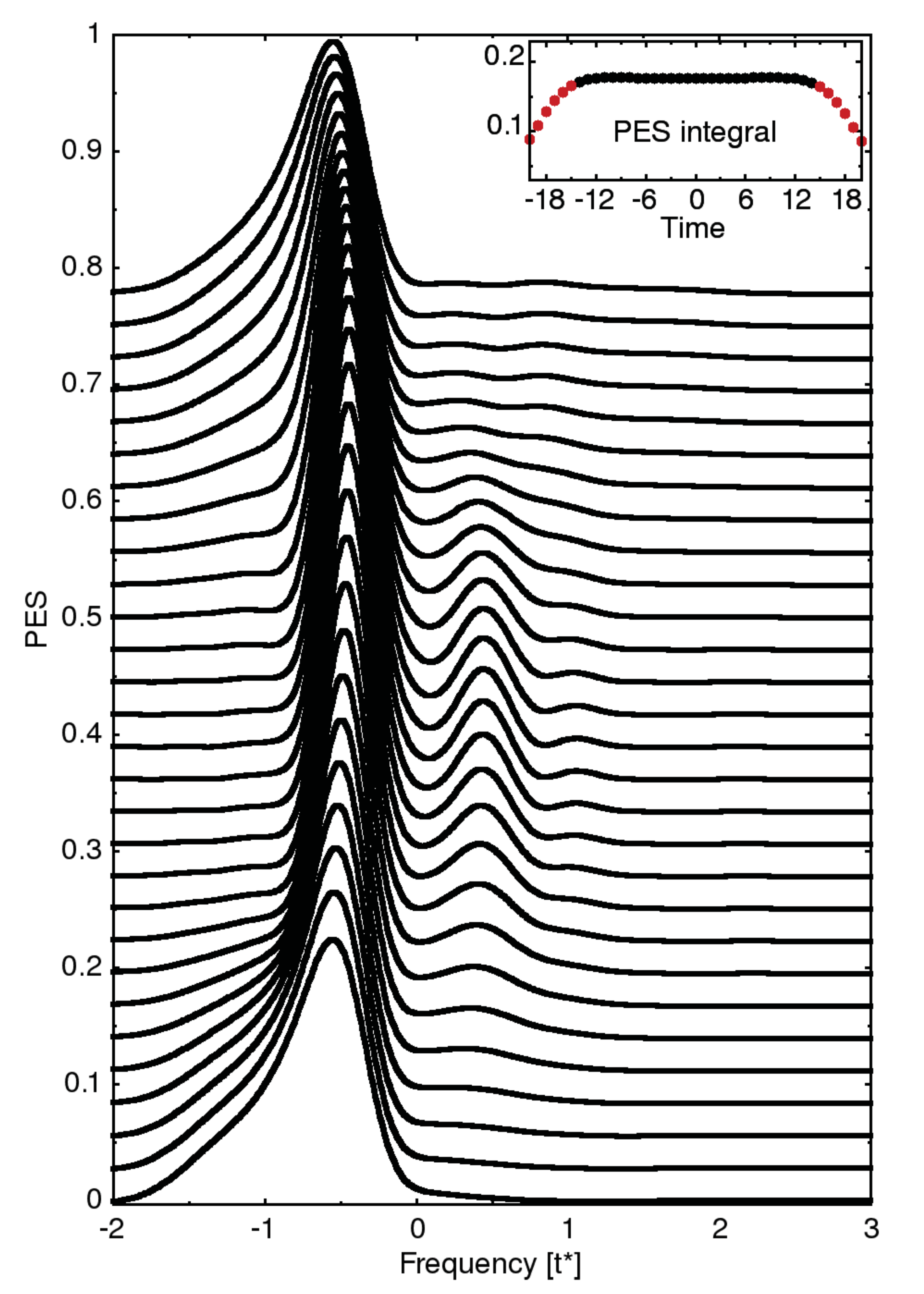}}
\caption{
(a--d, left) False-color plot of the calculated time-resolved PES for various $T$
with $E_0 = 30$, $U=0.86$ and averaged over the A and B sublattices,
plotted in false color. This system is a strongly correlated metal. The electric field is shown above the
plot.   The different panels are for the different temperatures in Fig.~\ref{fig: cdw_dos_fk_crit} with (a) the lowest and (d) the highest temperature.  (right) Waterfall image of vertical cuts through the TR-PES data plotted for different delay times and
offset for clarity. Inset is the total integrated spectral weight. Only the data that conserves the spectral weight is shown. This data corresponds to the case of panel (a) to the left.
[Reprinted and adapted from \cite{cdw_prb}, with permission]}
\label{fig: pes_fk_crit}
\end{figure}

As we move to Fig.~\ref{fig: pes_fk_crit}, we find even more surprising results. Here, it is virtually impossible to excite the electrons at all into the upper band. At the lowest temperature, the number excited is almost equal to the number de-excited. We continue to have the same band narrowing and peak sharpening due to field dressing of the states and the reduction of the spectral band gap for the CDW as before. But now, the real hallmark is that the number of electrons excited to the upper band is small after the pump pulse is completed. This compares reasonably to what we saw earlier for the order parameter, where we saw that it did not change significantly for the quantum critical CDW. Perhaps this behavior is tied to the existence of the conducting channel at the chemical potential in equilibrium; if that conducting channel is efficient, it can lead to a channel for de-excitation which competes with the excitation and leads to a small net excitation.

\begin{figure}[htb]
\centerline{
\includegraphics[width=0.49\textwidth]{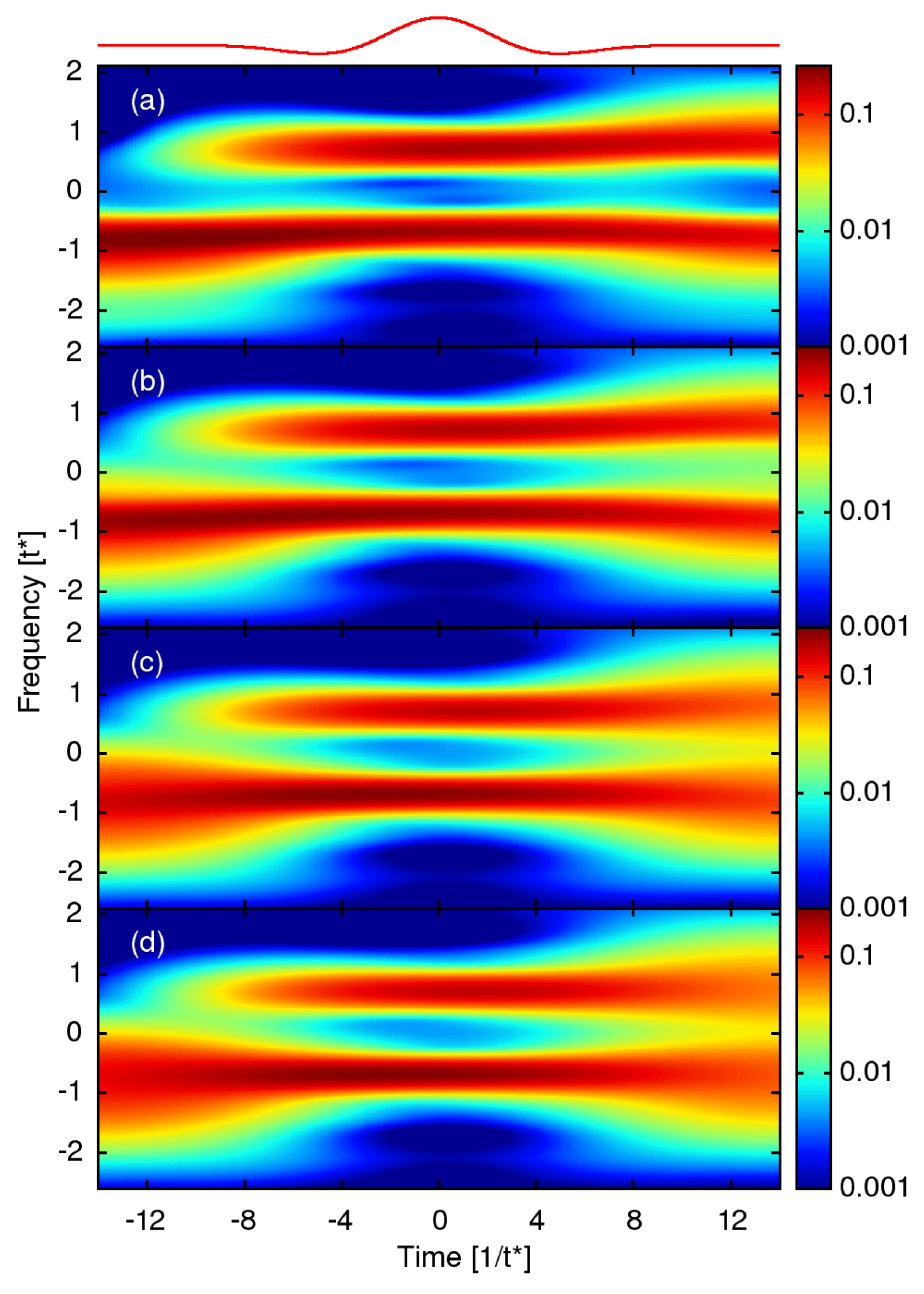}
\includegraphics[width=0.49\textwidth]{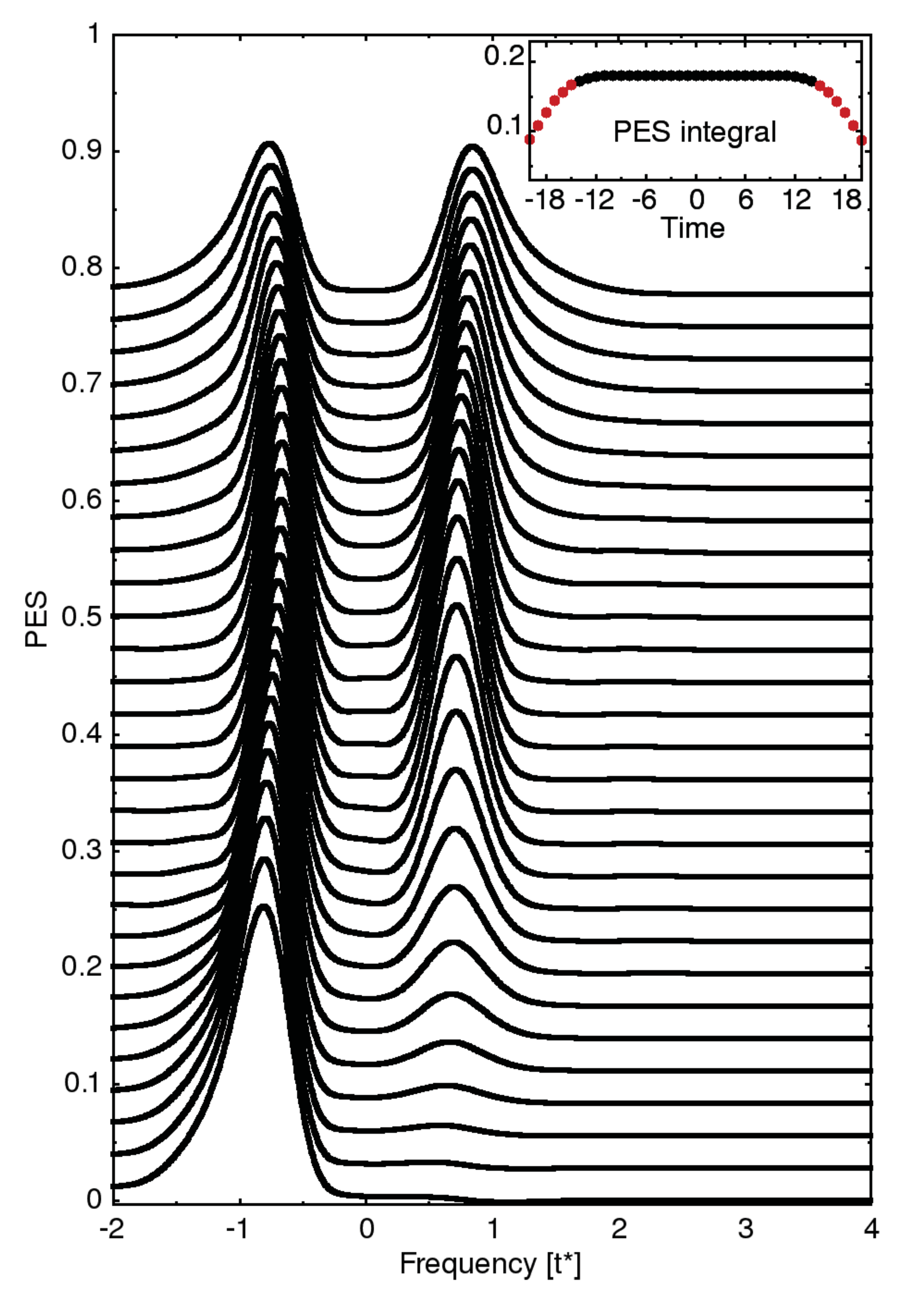}}
\caption{
(a--d, left) False-color plot of the calculated time-resolved PES for various $T$
with $E_0 = 30$, $U=1.4$ and averaged over the A and B sublattices,
plotted in false color. This system is a strongly correlated metal. The electric field is shown above the
plot.   The different panels are for the different temperatures in Fig.~\ref{fig: cdw_dos_fk_mott} with (a) the lowest and (d) the highest temperature.  (right) Waterfall image of vertical cuts through the TR-PES data plotted for different delay times and
offset for clarity. Inset is the total integrated spectral weight. Only the data that conserves the spectral weight is shown. This data corresponds to the case of panel (a) to the left.
[Reprinted and adapted from \cite{cdw_prb}, with permission]}
\label{fig: pes_fk_mott}
\end{figure}

In Fig.~\ref{fig: pes_fk_mott}, we show the same results for the critical Mott insulator. Here, we see almost complete excitation in the system---the weight in the upper and the lower bands appears to be nearly equal. We continue to see the band narrowing and the spectral gap narrowing, which we saw before, but now, this system has quite strong excitation. This result is surprising, because it excites more than the metal does. Perhaps, in this case, the de-excitation pathway suffers from some bottlenecks which make it less efficient. In any case, the other curious feature is that the gap closure due to subgap states is much more modest here, and in fact, it closes with the initial and final fields, but reopens when the field amplitude is the largest! A truly surprising result.

In experiment, the most important features seen are the closing of a gap by filling in of subgap states, its reforming well after the pump, along with an oscillation of the PES that modulates at the same frequency as the phonon that is responsible for the CDW order. In addition, for the TbTe$_3$ experiment, as the fluence is turned up, the system has the spectral gap get reduced, but it never goes all the way to zero. Some of these features are captured in these simplified models of a CDW while others are not. What these all electronic models see is a gap collapse by filling of subgap states, and a partial reduction of the spectral gap, but not all the way to zero. They do not see the collapse of the gap for long times after the pump, nor do they see the long-time oscillations, both which are likely associated with the ordering phonon (which is not part of this model). The model calculations also show a decoupling of what is happening with the CDW order parameter, as measured by the modulation of the charge, and the spectral gaps, as measured in a PES experiment.

The model calculations also see some new phenomena which is yet to be seen in experiment. This includes the preponderence of de-excitation and the difficulty with exciting a quantum critical CDW. It also includes the large band narrowing and peak sharpening due to field dressing when the pump field is on. 

\section{Conclusions}

What is next? From the theory standpoint, the next step is to include the coupling to the phonons directly. This should reproduce a number of the additional features seen in experiment, and will hopefully allow for a more complete solution of the problem. From the experimental standpoint, we hope that experiments might be done on materials that have there CDW transition driven by nesting, where these additional subgap features and quantum critical behavior may be present. Those systems could prove to be an interesting playground for novel physics.

In this review, we covered a wide range of the theory needed to describe strongly correlated CDW materials in nonequilibrium. We developed the theory both for the simplest model of a CDW, which is essentially a bandstructure with a basis, to describing the exact solution of an all electronic CDW described by the Falicov-Kimball model. The latter having a unique quantum critical point within the CDW phase. We used these formalisms to solve these problems for a wide array of different parameters. Technical reasons restricted the Falicov-Kimball model calculations to large amplitude fields only. 

This work represents starting point for the theory of nonequilibrium pump/probe experiments in strongly correlated materials. Much more work needs to be done in the future to treat different kinds of order, the competition between different ordered phases, and different types of electron correlations. In addition, extensions that include the possibility of incorporating more real materials properties, along the lines of density functional theory plus dynamical mean-field theory, but now in nonequilibrium, will also be important. We hope to see and to participate in these developments in the coming years.

\ack
This work was supported by the Department of Energy, Office of Basic Energy Sciences, Division of Materials Sciences and Engineering under Contract Nos. DE-AC02-76SF00515 
(Stanford/SIMES) and DE-FG02-08ER46542 (Georgetown). Computational resources were provided by the National Energy Research Scientific 
Computing Center supported by the Department of Energy, Office of Science, under Contract No. DE-AC02-05CH11231. J.K.F. was also supported by the McDevitt Bequest at Georgetown.

\section{Referencing\label{except}}

\end{document}